\documentclass[useAMS,usenatbib,usegraphicx]{mn2e}

\title[The Kormendy relation at $z\sim$1.5]
{The Kormendy relation of massive elliptical galaxies at $z\sim$1.5. Evidence
for size evolution ?}
\author[M. Longhetti et al.]{M. Longhetti$^{1}$, 
P. Saracco$^{1}$\thanks{E-mail: paolo.saracco@brera.inaf.it}, 
P. Severgnini$^{1}$, R. Della Ceca$^{1}$,
\newauthor F. Mannucci$^{2}$, R. Bender$^{3,4}$,
N. Drory$^{4,5}$, G. Feulner$^{3,4,6}$, U. Hopp$^{3,4}$ \\
$^{1}$INAF - Osservatorio Astronomico di Brera, Via Brera 28, 20121 Milano\\
$^{2}$INAF - Istituto di Radioastronomia, Largo E. Fermi 5, 50125 Firenze, 
Italy\\
$^{3}$Universit\"ats-Sternwarte M\"unchen, Scheiner Str. 1, D-81679
M\"unchen, Germany\\
$^{4}$Max-Planck-Institut f\"ur extraterrestrische Physik,
Giessenbachstra\ss e, D-85748 Garching, Germany\\
$^{5}$University of Texas at Austin, Austin, Texas 78712, USA\\
$^6$ Potsdam--Institut f\"ur Klimafolgenforschung, Postfach 
  60~12~03, D-14412 Potsdam, Germany
}
\begin{document}

\date{Accepted 2006 October 09. Received 2006 September 15; in original form 2006 July 27}
\pagerange{\pageref{firstpage}--\pageref{lastpage}} \pubyear{2006}
\maketitle
\label{firstpage}
\begin{abstract}
We present the morphological analysis based on HST-NIC2 (0.075 arcsec/pixel) 
images in the F160W filter of a sample of 9 massive field 
($> 10^{11} $ 
M$_\odot$) galaxies spectroscopically classified as early-types at $1.2<z<1.7$.
Our analysis shows that all of them are bulge dominated systems.
In particular, 6 of them are  well fitted by a  de Vaucouleurs profile 
($n=4$) 
suggesting that they can be considered pure elliptical galaxies. 
The remaining 3 galaxies are better fitted by a S\'ersic profile with index
$1.9<n_{fit}<2.3$ suggesting that a disk-like component could contribute up 
to 30\% to the total light of these galaxies.
We derived the effective radius R$_e$ and the mean surface brightness 
$\langle\mu\rangle_e$ within R$_e$ of our galaxies and we compared them with 
those of early-types at lower redshifts.
We find that the surface brightness $\langle\mu\rangle_e$ of our galaxies 
should get fainter by 2.5 mag from $z\sim1.5$ to $z\sim0$ to match the surface 
brightness of the local ellipticals with comparable R$_e$, 
i.e. the local Kormendy relation.
Luminosity evolution without morphological changes can only explain
half of this effect, as the maximum dimming expected for an elliptical
galaxy is $\sim$1.6 mag in this redshift range.
Thus, other parameters, possibly structural, may undergo  evolution 
and play an important role in reconciling models and observations.
Hypothesizing an evolution of the effective radius of galaxies 
we find that R$_e$ should increase by a factor 1.5 from $z\sim1.5$ 
to $z\sim0$.
\end{abstract}

\section{Introduction}
The working out of the hierarchical clustering paradigm paved the way to the 
development of various models of galaxy formation.
All these models are based on the assumption that gravity is the main driver
shaping the structures, both on galactic and larger scales.
Observational evidences sustaining this scenario, in which  structures 
form by subsequent mergers of dark matter haloes, 
come from the WMAP data which constrain the fundamental 
parameters of modern cosmology ($\Omega_0$, the Hubble constant and the 
amount of dark energy) to better then a few per cent (e.g. Spergel et al. 
2003).
However, while the general build-up of  cosmic structures seems
to be well described, the assembly of the baryonic mass on galactic scales
still represents a weak point and it does still not find a satisfactory 
description in these galaxy formation models. 
In particular, the assembly of high-mass spheroids is a key issue of the 
current models of galaxy formation since they still miss the process 
accounting for their growth.

As recently pointed out by Nagamine et al. (2005a), one of the difficulties 
of hierarchical models consists indeed in accounting for the population of 
massive (M$_{star}>10^{11}$ M$_\odot$) early-type galaxies seen at high $z$.
Indeed, while they show that high-mass optically selected (blue) galaxies at
$z\simeq3$, i.e. the Ly-break galaxies, can be easily recovered in the
framework of hierarchical models (Nagamine et al. 2005b),
the number of massive early-type (red and dead) galaxies seen at $z>1$
hardly fit the predictions of these  models
(see also Somerville et al. 2004).
A recent important result  has been, indeed, the discovery and the 
spectroscopic identification of a significant population of massive 
early-type galaxies at $1.2<z<1.7$ (McCarthy et al. 2004; 
{ Glazebrook et al. 2004;} Saracco et al. 2005).
Their number density  is comparable to the number density of massive 
early-types seen in the local Universe showing that no strong evolution 
has taken place up to $z\simeq1.5$ (Saracco et al. 2005).
On the other hand,  their spectrophotometric properties
reveal a spread in the mean age of their stellar population suggesting
a different star formation history and/or a different redshift of formation
(Longhetti et al. 2005).
The discovery of  early-type galaxies at $z\simeq2$
(Cimatti et al. 2004; Daddi et al. 2005) shows that early-types exist at 
least up to $z\simeq2$.

Another important observational evidence is the mass-dependent evolution 
which the galaxies undergo apparently (Gavazzi 1993; Gavazzi et al. 1996), 
the so called ``downsizing'' (Cowie at al. 1996).
This observed behaviour constrains the build up of the stellar mass in 
high-mass early-type galaxies earlier and over a shorter interval than in 
the lower mass ones.
This behaviour has been recently confirmed observationally both at low and 
at high redshift through a variety of analysis: 
from the abundance ratios of local early-type galaxies (Thomas et al. 2005),
{ from the evolution of the fundamental plane (Treu et al.2005)}, 
of the specific star formation rate 
(SSFR; Feulner et al. 2005) and of the stellar mass function 
(Drory et al. 2005; { Bundy et al. 2005}) of galaxies,  
from the clustering of distant red galaxies 
(Grazian et al. 2006), from  the optical LF of galaxies at $z<1$ 
(Cimatti et al. 2006) and from the evolution of the near-IR LF of galaxies to
$z\sim3$ (Saracco et al. 2006; Caputi et al. 2006).

These pieces of evidence are apparently at variance with the expectations of
hierarchical models which predict a $z>2$ Universe mostly populated  by 
star-forming disks, irregular galaxies and merging systems and in which 
the local massive spheroids complete their assembly last, possibly at $z<1$. 

To further complicate this picture, evidence for a possible lack of big
early-type galaxies at $z\sim1$ and beyond are coming out
(e.g. Fasano et al. 1998; 
Waddington et al. 2002; Daddi et al. 2005; Trujillo et al. 2006a, 2006b; 
Cassata et al. 2005).
At first glance, this could be interpreted as a prove of the merging process
responsible of the growth of local high-mass ellipticals.
On the other hand, the apparently small early-types seen at high-redshift 
seem to be characterized by an effective surface brightness brighter than  
the one of local counterparts even considering  luminosity evolution.
Moreover, their stellar masses exceed those of local galaxies with comparable
size.
Thus, they seem to be more compact then their local counterparts.
It should be noted, however, that most of these results are based on HST
optical observations sampling the blue and UV rest-frame emission of the 
galaxies, particularly sensitive to morphological k-correction and star 
formation episodes, and/or on seeing limited ground-based observations.
These features could affect the estimate of the effective radius
of high-z early-type galaxies.

In this paper, we present the morphological analysis based on high-resolution
(FWHM$\sim$0.1 arcsec) HST-NICMOS observations in the F160W filter 
($\lambda\sim1.6$ $\mu$m) of a sample of 9 high-mass early-type galaxies
with spectroscopic confirmation at $1.2<z<1.7$.
The filter used samples the rest-frame R-band at the redshift of our galaxies. 
The aim of this work is to establish whether the surface brightness vs. 
effective radius relation of early-type galaxies at $z>1$ can be accounted for
by their luminosity evolution or whether they are in fact more compact 
than local 
early-types without the uncertainties related to the optical and
ground based observations.

The paper is organized as follows:
in \S 2 we describe the sample and the HST observations.
In \S 3 we describe the profile fitting procedure used, we derive 
the morphological parameters of our sample of early-types and we
describe the simulations  performed to assess
the robustness of our fitting.
In \S 4 we report the results
of our morphological analysis.
In \S 5 we discuss the Kormendy relation, we compare our results with those 
obtained at lower redshift by other authors, we constrain the evolution that 
early-types must undergo at $z<1.5$ and we propose a model accounting for 
the observed evolution.
Summary and conclusions are given in \S 6.

Throughout this paper we assume Vega magnitudes, H$_0$=70 km s$^{-1}$
Mpc$^{-1}$, $\Omega_0=0.3$ and $\Lambda_0=0.7$ flat cosmology.

\section{Sample and HST-NICMOS observations}
The sample is composed of 9 out of the 10 field galaxies spectroscopically 
classified as early-types at  $1.2<z<1.7$ ($\langle z\rangle=1.4$, 
Longhetti et al. 2005).
They  result from a near-IR spectroscopic follow-up of a complete sample of 
bright (K$<18.5$) EROs (R-K$>5.3$)  selected from the Munich Near-IR 
Cluster Survey (MUNICS, Drory et al. 2001) that provides 
optical  (B, V, R, I) and near IR (J and K') photometry.
The low resolution spectroscopic observations were carried out with 
the prism disperser AMICI at the near-IR camera NICS of the TNG 
(Telescopio Nazionale Galileo, La Palma, Canary Islands).
The spectroscopic and the photometric data show that
they have stellar masses greater than $10^{11} $ M$_\odot$ and that
they account  for more than 70\% of the number and the stellar mass density 
of local early-type galaxies with comparable masses (Saracco et al. 2005).
The stellar population is  3-5 Gyr old in
6 of them and  1-2 Gyr old in the remaining 4 galaxies implying  minimum 
formation redshifts $z_f>4$   and $z_f>2$ respectively (Longhetti et al. 2005).
The properties of the 9 early-type galaxies (absolute K-band magnitude,
stellar mass and age) are summarized in Tab.1.
\noindent

\begin{table}
\caption{Properties of the massive early-type galaxies of our 
sample. Details on all the
reported parameters can be found in Longhetti et al. (2005). 
}
\centerline{
\begin{tabular}{lcccc}
\hline
\hline
  Object  & $z_{spec}$ & M$_{K}$ & $\mathcal{M}_{star}$ & $age_w$  \\
          &     &         & [10$^{11}$ M$_{\odot}$] & [Gyr] \\
  \hline
{S2F5\_109} & 1.22$\pm$0.05 & -27.3$\pm$0.1 & 4.7 - 14.6 & 1.7$\pm$0.3  \\
{S7F5\_254} & 1.22$\pm$0.05 & -26.3$\pm$0.1 & 4.5 - \ 9.0 & 5.0$\pm$0.1  \\
S2F1\_357 & 1.34$\pm$0.05 & -26.4$\pm$0.1 & 4.2 - \ 9.4 & 4.0$\pm$0.1   \\
S2F1\_389 & 1.40$\pm$0.05 & -26.3$\pm$0.2 & 1.7 - \ 6.5 & 3.0$\pm$0.5  \\
{S2F1\_511} & 1.40$\pm$0.05 & -26.1$\pm$0.1 & 1.1 - \ 5.5 & 1.3$\pm$0.3   \\
{S2F1\_142} & 1.43$\pm$0.05 & -26.5$\pm$0.1 & 3.1 - \ 9.3 & 2.2$\pm$0.2   \\
S7F5\_45\  & 1.45$\pm$0.05 & -26.8$\pm$0.2 & 2.3 - 11.2 & 1.7$\pm$0.3 \\
S2F1\_633 & 1.45$\pm$0.05 & -26.2$\pm$0.1 & 3.1 - \ 7.5 & 4.0$\pm$0.5  \\
S2F1\_443 & 1.70$\pm$0.05 & -26.6$\pm$0.2 & 2.0 - \ 9.4 & 3.5$\pm$0.3 \\
\hline
\hline
\end{tabular}
}
\end{table}

The near-IR HST observations were carried out during Cycle 14
with the NICMOS-NIC2  camera (0.075 arcsec/pixel) in the F160W filter 
($\lambda\sim16030$ \AA).
Considering the redshift of our galaxies, the selected filter samples their 
R-band rest-frame continuum.
Each galaxy has been observed for 2 orbits ($\sim4800$ seconds exposure) 
with the exception of the brightest one (S2F5\_109) for which we allocated a 
single orbit (2400 sec).
The observations consist of  8 dithered exposures of about 300 seconds each
for each orbit.
The MULTIACCUM Read Out Mode STEP16 with NSAMP=25 and a dither size 
of 4 arcsec with a spiral dither pattern have been adopted.

The HST calibrated images used for our analysis are those provided by 
the HST data archive resulting from the standard automatic reduction 
procedures CALNICA and CALNICB.
The  limiting surface brightness reached in our images
is $\mu_{F160}\simeq27.77$ mag/pix (29.11 AB mag/pix) corresponding to
$\mu_{F160}\simeq22.14$ mag/arcsec$^2$ (23.48 AB mag/arcsec$^2$).
This allows us to sample the light profile of our galaxies at a  S/N$>$3 
out to more than 3 effective radii. 


\section{Morphological parameters}
The measure of the morphological parameters of the galaxies,
such as their effective radius ($r_{e}$ in arcsec or R$_e$ in kpc) 
and the mean effective surface 
brightness within R$_e$ ($\langle\mu\rangle_{e}$ in mag/arcsec$^{2}$),  
has been performed by means of the \texttt{Galfit} software (v. 2.0.3; 
Peng et al. 2002).
\texttt{Galfit} builds a bi-dimensional model of the image of the galaxies
on the basis of an analytic function that can be chosen by the user.
It takes into account the position angle (PA) of the major axis $a$ of the 
profile and the ratio $a/b$ between major and minor axis.
The bi-dimensional model is convolved with the 
Point Spread Function (PSF) of the observations before being compared with 
the observed image.
Minimization of the differences between models and observations
gives the best fitting parameters which have
been left varying in the procedure. 

We  adopted a single component to fit the light distribution of the galaxies 
in our sample.
We performed the fits adopting in separate steps the de Vaucouleurs (1948),
the exponential and the S{\'e}rsic (1968) profiles.
The analytic expression of the adopted profiles is

\begin{equation}
I(r)=I_{e} exp\{-b_{n}[(r/r_{e})^{1/n}-1]\}
\end{equation}

\noindent
where $I(r)$ is the surface brightness measured at the distance
$r$ from the centre of the galaxy in flux units/arcsec$^{2}$, 
$I_{e}$ is the surface brightness
measured at $r=r_{e}$ where $r_{e}$ [arcsec] 
and $b_{n}$ is a normalisation factor depending
on the exponent {\it n} chosen for the fit.
The exponent {\it n} is a free parameter for the S{\'e}rsic profile,
while it is  $n=1$ and $n=4$ for the exponential and de Vaucouleurs
profiles respectively.
The other free parameters in the adopted fits are
the total magnitude (mag$_{Gal}$), the semi-major axis $a_{e}$ of the
projected elliptical isophote containing half of the total light,
the axial ratio $b/a$ and the position angle of the major axis.
The circularised effective radius $r_{e}$ is related to $a_{e}$
through the equation $r_{e}=a_{e} \sqrt{b/a}$.
The mean effective surface brightness $\langle\mu\rangle_{e}$ can be computed
considering half of the total flux of the galaxy and the area
included within the effective radius
\begin{equation} 
\langle\mu\rangle_{e}=F160W(mag)+5log(r_{e})+2.5log(2\pi)
\end{equation}
{ The magnitude in the F160W filter used to derive $\langle\mu\rangle_{e}$
is the total magnitude mag$_{Gal}$ derived by \texttt{Galfit}}.

A different PSF has been generated  for each galaxy by means
of the \texttt{Tiny Tim}\footnote{www.stsci.edu/software/tinytim}
(v. 6.3) software package (Krist 1995; Krist J. \& Hook R. 2004).
Since the final image of each galaxy is the result of the co-adding 
of 8 shifted images, the relevant PSF has been constructed by  averaging 
the 8 PSFs generated at each target position. 
No sub-sampling has been adopted
and the best fitting Spectral Energy Distribution (SED) of each
galaxy (Saracco et al. 2003, 2005; Longhetti et al. 2005) has been
used to simulate the spectral shape of the light entering
the F160W filter. 
We  verified that the results of the fitting do not depend on the 
specific choice of the \texttt{Tiny Tim} parameters mentioned above. 
In particular, we find that the best fitting parameters  vary less than 
2\% (e.g. $\Delta (r_e)<0.01''$) for different choices of SED, 
sub-sampling factors and  PSF position.

\begin{figure*}
\vskip -2truecm
\centering
\includegraphics[width=18.5cm,height=21.5cm]{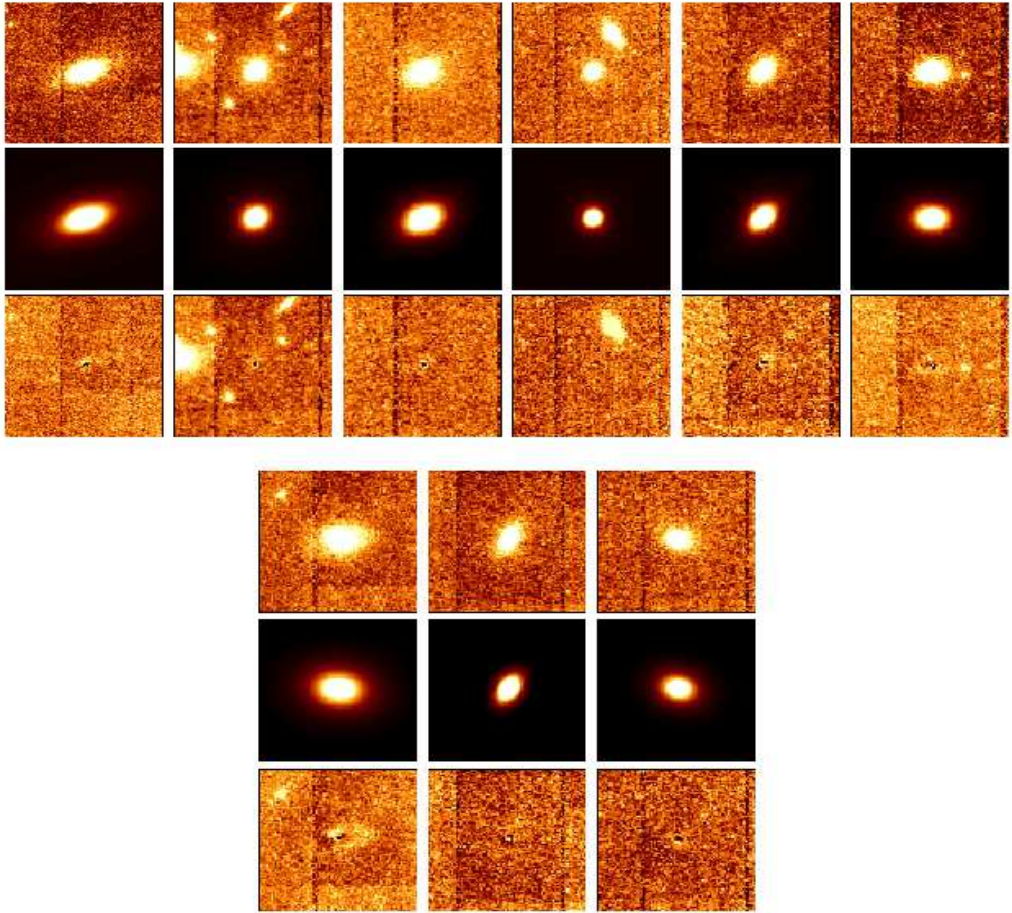}
\vskip -4truecm
\caption{Each column shows the \texttt{Galfit} input and output for 
the 9 early-type galaxies: NIC2 F160W-band image of the galaxy (upper panel),
best-fitting de Vaucouleurs model profile (middle panel) and residual
image (lower panel) obtained by subtracting the model from the image.
Galaxies are (from top-left to bottom-right): S2F5\_109, S7F5\_254,
 S2F1\_357, S2F1\_389, S2F1\_511, S2F1\_142, S7F5\_45, S2F1\_633, S2F1\_443. 
Each image is 6$\times6$ arcsec.
}
\end{figure*}

\subsection{Profile fitting}
We first assess whether the light distribution of our galaxies
is bulge-dominated, as expected from our spectroscopic classification,
or disk-dominated.
To this end we compared the goodness of fit obtained with the 
de Vaucouleurs ($n=4$, bulge) profile and the exponential ($n=1$, disk) 
one making use of the relevant $\chi^2$.
The fitting has been performed over a box of 6$\times6$ arcsec
 centered on the centroid of 
each galaxy after having checked for the stability of the results 
with respect to the size of the box.
As initial guesses for the centroid, for the PA and for the 
total magnitude of the galaxies we used the SExtractor  output.
{ We did not keep the SExtractor MAG\_BEST value as the actual total 
magnitude of the galaxies since it turned out to be less accurate than the one 
resulting from \texttt{Galfit}, as we verified through simulations 
(see next sub-section).}
Neighbouring sources and bad columns resulting from the mosaicking have
been masked out from the model fitting.
The root mean square image produced by the standard HST-NICMOS reduction 
procedure has been used to give the relative weights to the sky pixels in
the fitting. 
We find that the de Vaucouleurs profile  always provides a better fitting  
than the exponential  one for all the galaxies.
In particular, the $n=1$ profile provides a fitting significantly worse 
($\Delta(\chi^2_\nu)>0.3$,  d.o.f.$\gg100$) than the de Vaucouleurs one for  
8 early-type galaxies while, for galaxy S2F1\_443, the difference 
is not significant ($\Delta(\chi^2_\nu)\sim0.01$). 
It is interesting to note that S2F1\_443 is the highest redshift galaxy of 
our spectroscopic sample of early-types.
Thus, both the surface brightness dimming and the angular scale are larger 
than for the others.
Besides this,  this galaxy is ''peculiar'' with respect 
to the other galaxies  since it is  X-ray emitting,  
probably hosting a type 2 QSO (Severgnini et al. 2005). 

From this first comparison, we can exclude that a disk component is the 
dominant feature of the light distribution of our  early-type galaxies, 
in agreement with their spectrophotometric properties. 
In Fig. 1  the F160W-band image, the best-fitting 
de Vaucouleurs model and the residual of the fitting are shown for each galaxy.

We then compared the Sersic profile to the de Vaucouleurs profile 
to verify whether the addition of the Sersic index $n$ as free parameter 
improves significantly the fitting and to find the best fitting structural 
parameters.  
The best fitting Sersic index $n$ obtained for 6 out of the 9 galaxies
is in the range $2.8<n<4.5$.
This result confirms that all of them are 
dominated by a bulge component.
It is worth noting that the addition of the free parameter $n$ does not 
significantly improve  the fitting of these 6 galaxies with respect to the 
de Vaucouleurs profile, i.e. it is not statistically required as
shown by the reduced $\chi^2$ we obtained.
Thus, we can consider these galaxies morphologically as pure elliptical 
galaxies.
For the remaining 3 galaxies, S7F5\_254, S7F5\_45 and S2F1\_443, 
a Sersic index $n<2.5$  provides a better fitting to the data suggesting
that a non negligible fraction, up to 30\% of the light can be 
distributed on a disk component 
(see Saglia et al. 1997; van Dokkum et al. 1998).
Given the spectrophotometric early-type nature of these galaxies
(see Longhetti et al. 2005), we conclude that they probably are  S0/Sa 
galaxies. 

Table 2 reports, for each galaxy, the total magnitude in the F160W filter and 
the morphological parameters we derived assuming the de Vaucouleurs and 
the S{\'e}rsic  profiles. 
{ The reduced $\chi^2$ relevant to each best-fitting profile is also reported.
The values of the effective radii (r$_e$ and R$_e$) are corrected for the small
underestimate we found from our simulations (see next sub-section).
The mean surface brightnesses $\langle\mu\rangle_e^{F160W}$ and 
$\langle\mu\rangle_e^R$) have been derived from the corrected radii.
}

\begin{table*}
\caption{Morphological parameters of the galaxies.
For each galaxy the first line reports the results obtained with the 
S{\'e}rsic profile while the second line reports data obtained with
$n=4$ (de Vaucouleurs profile). Magnitudes are in the Vega system.
F160W magnitudes can be transformed into AB mag by adding 1.36.
{ The effective radii obtained with the Sersic and the de Vaucouleurs
profiles have been corrected by about 0.07'' and 0.025'' respectively, 
according to the results of our simulations (see \S 3.2).
The surface brightnesses have been derived from the corrected radii}.   
}
\centerline{
\begin{tabular}{lclcccccc}
\hline
\hline
Object       & F160W  &{\it n} & r$_e$  & R$_{e}$ & $\langle\mu\rangle_{e}^{F160W}$ & $\langle\mu\rangle_{e}^{R}$    & {\it b/a} & $\chi^2_{\nu}$  \\
             & [mag] &        & [arcsec] &[kpc]   & [mag/arcsec$^{2}$]   & [mag/arcsec$^{2}$] & &         \\
\hline
 S2F5\_109   &    17.47$\pm0.03$ & 3.0$\pm0.04$ & 0.53$\pm0.01$ & 4.43$\pm0.07$	& 18.10$\pm0.05$ & 20.11$\pm0.05$ & 0.50$\pm0.01$ & 1.34\\ 
  	     &    17.28$\pm0.03$ & 4.0	 	& 0.67$\pm0.01$ & 5.57$\pm0.07$	& 18.41$\pm0.04$ & 20.42$\pm0.04$  & 0.49$\pm0.01$ & 1.36\\
 S7F5\_254   &    19.46$\pm0.03$ & 2.3$\pm0.10$	& 0.27$\pm0.01$ & 2.28$\pm0.11$	& 18.65$\pm0.13$ & 20.77$\pm0.13$ & 0.82$\pm0.02$ & 2.47\\
  	     &    19.20$\pm0.03$ & 4.0	 	& 0.34$\pm0.01$ & 2.80$\pm0.11$	& 18.84$\pm0.09$ & 20.96$\pm0.09$ &  0.82$\pm0.02$ & 2.50\\
 S2F1\_357   &    18.72$\pm0.03$ & 2.8$\pm0.08$	& 0.33$\pm0.01$ & 2.78$\pm0.07$	& 18.31$\pm0.07$ & 20.41$\pm0.07$ &  0.67$\pm0.01$ & 0.95\\
  	     &    18.53$\pm0.03$ & 4.0	 	& 0.39$\pm0.01$ & 3.28$\pm0.07$	& 18.48$\pm0.06$ & 20.58$\pm0.06$ &  0.66$\pm0.01$ & 0.96\\
 S2F1\_389   &    19.79$\pm0.03$ & 4.5$\pm0.40$	& 0.25$\pm0.02$ & 2.07$\pm0.18$	& 18.74$\pm0.26$ & 20.79$\pm0.26$ &  0.85$\pm0.03$ & 1.46\\
  	     &    19.85$\pm0.03$ & 4.0	 	& 0.18$\pm0.02$ & 1.54$\pm0.15$	& 18.16$\pm0.24$ & 20.21$\pm0.24$ &  0.86$\pm0.03$ & 1.46\\
 S2F1\_511   &    19.15$\pm0.03$ & 3.3$\pm0.13$	& 0.25$\pm0.01$ & 2.07$\pm0.07$	& 18.10$\pm0.11$ & 20.09$\pm0.11$ &  0.60$\pm0.01$ & 1.34\\
  	     &    19.07$\pm0.03$ & 4.0	 	& 0.23$\pm0.01$ & 1.91$\pm0.07$	& 17.85$\pm0.09$ & 19.84$\pm0.09$ &  0.59$\pm0.01$ & 1.35\\ 
 S2F1\_142   &    18.65$\pm0.03$ & 3.5$\pm0.10$	& 0.36$\pm0.01$ & 3.04$\pm0.12$	& 18.43$\pm0.11$ & 20.48$\pm0.11$ &  0.74$\pm0.01$ & 1.74\\
  	     &    18.59$\pm0.03$ & 4.0	 	& 0.35$\pm0.01$ & 2.95$\pm0.07$	& 18.31$\pm0.06$ & 20.36$\pm0.06$ &  0.73$\pm0.01$ & 1.74\\
 S7F5\_45\   &    18.83$\pm0.03$ & 2.0$\pm0.10$ & 0.55$\pm0.02$ & 4.61$\pm0.17$ & 19.51$\pm0.09$ & 21.57$\pm0.09$ &  0.71$\pm0.01$ & 1.19\\
             &    18.25$\pm0.03$ & 4.0  	& 1.13$\pm0.04$ & 9.53$\pm0.33$ & 20.51$\pm0.08$ & 22.57$\pm0.08$ &  0.69$\pm0.01$ & 1.21\\
 S2F1\_633   &    19.00$\pm0.03$ & 4.1$\pm0.20$	& 0.31$\pm0.01$ & 2.63$\pm0.11$	& 18.46$\pm0.11$ & 20.52$\pm0.11$ &  0.53$\pm0.01$ & 1.37\\
  	     &    19.00$\pm0.03$ & 4.0	 	& 0.26$\pm0.01$ & 2.23$\pm0.07$	& 18.10$\pm0.08$ & 20.16$\pm0.08$ &  0.53$\pm0.01$ & 1.37 \\
 S2F1\_443   &    19.44$\pm0.03$ & 1.9$\pm0.07$	& 0.40$\pm0.02$ & 3.35$\pm0.15$	& 19.42$\pm0.12$ & 21.43$\pm0.12$ &  0.79$\pm0.02$ & 1.10\\
  	"    &    18.94$\pm0.03$ & 4.0	 	& 0.72$\pm0.03$ & 6.13$\pm0.24$	& 20.24$\pm0.09$ & 22.25$\pm0.09$ &  0.76$\pm0.02$ & 1.13\\
\hline
\hline
\end{tabular}
}
\vskip 5truecm
\end{table*}
Fig. 2 presents the
surface brightness (SB) profile of the 9 galaxies in the F160W band. 
The observed profile (filled points) extracted along the galaxy major axis
is compared to the best-fitting de Vaucouleurs (solid/red line) and  Sersic 
(dashed/green line) model profiles.
The \texttt{ellipse} task within the IRAF package has been used
to extract both the profiles on real and model images,
adopting the best fitting parameters (ellipticity and orientation)
provided by \texttt{Galfit}. 

\begin{figure*}
\includegraphics[width=5.8cm,height=5.8cm]{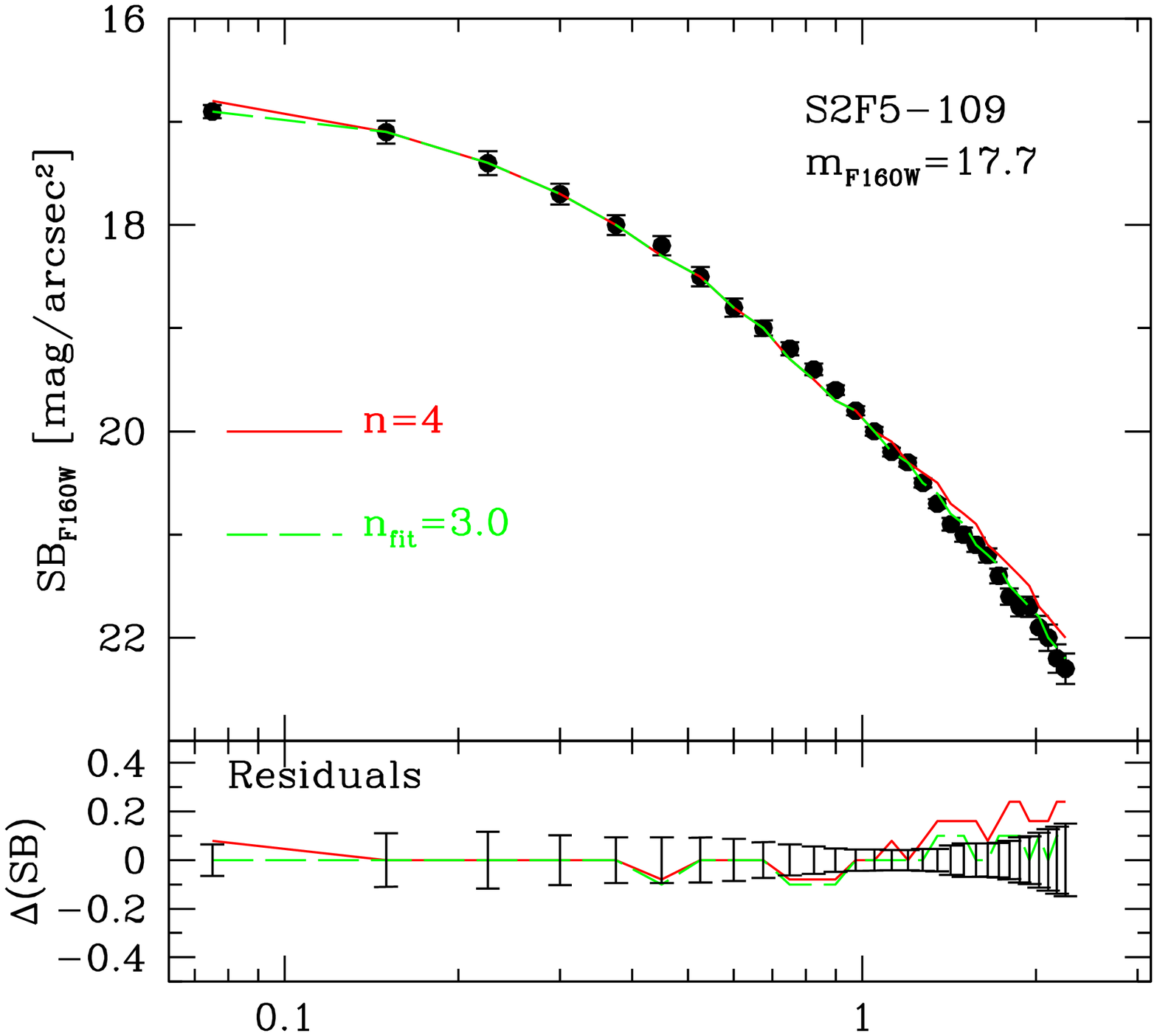}
\includegraphics[width=5.8cm,height=5.8cm]{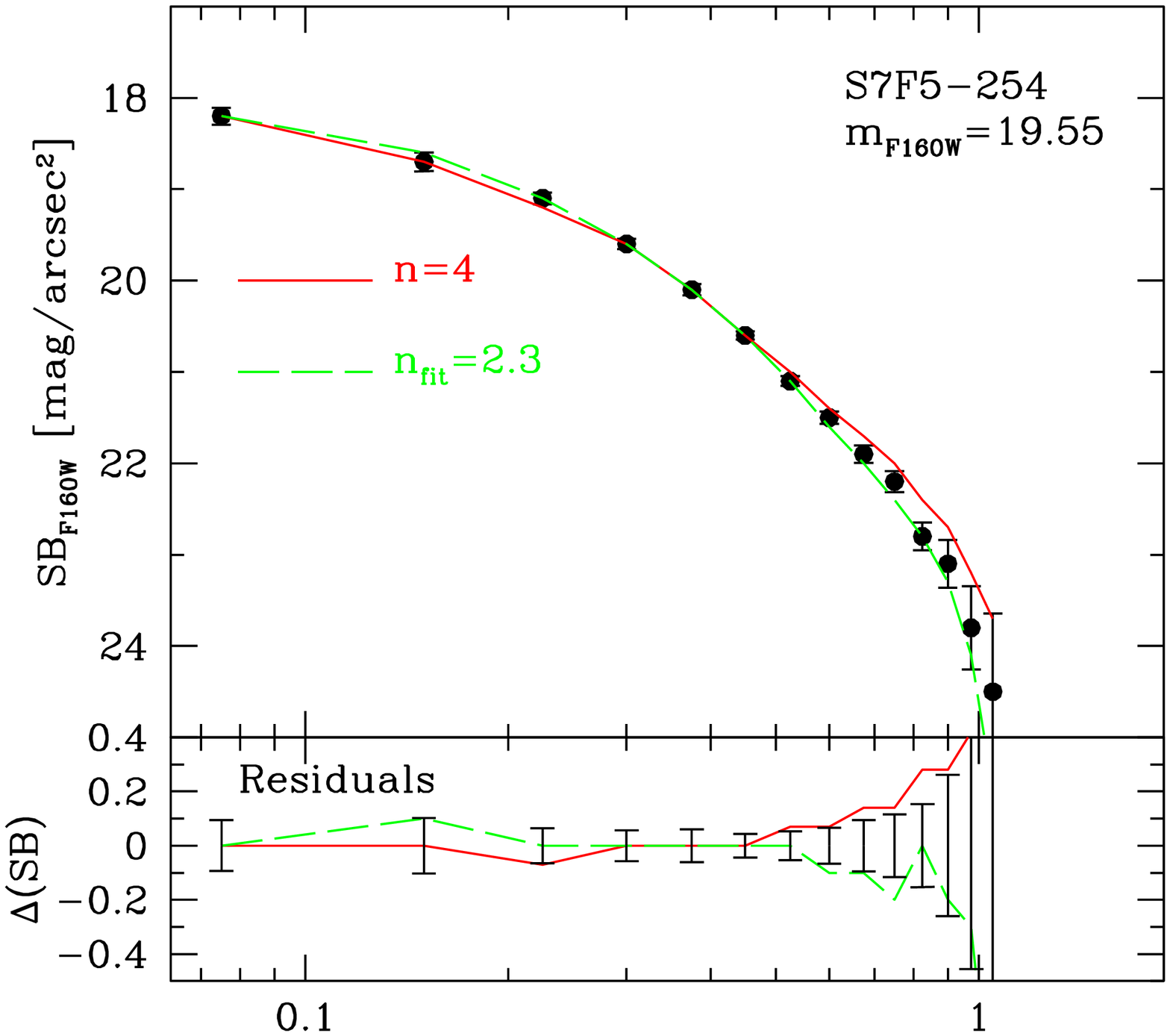}
\includegraphics[width=5.8cm,height=5.8cm]{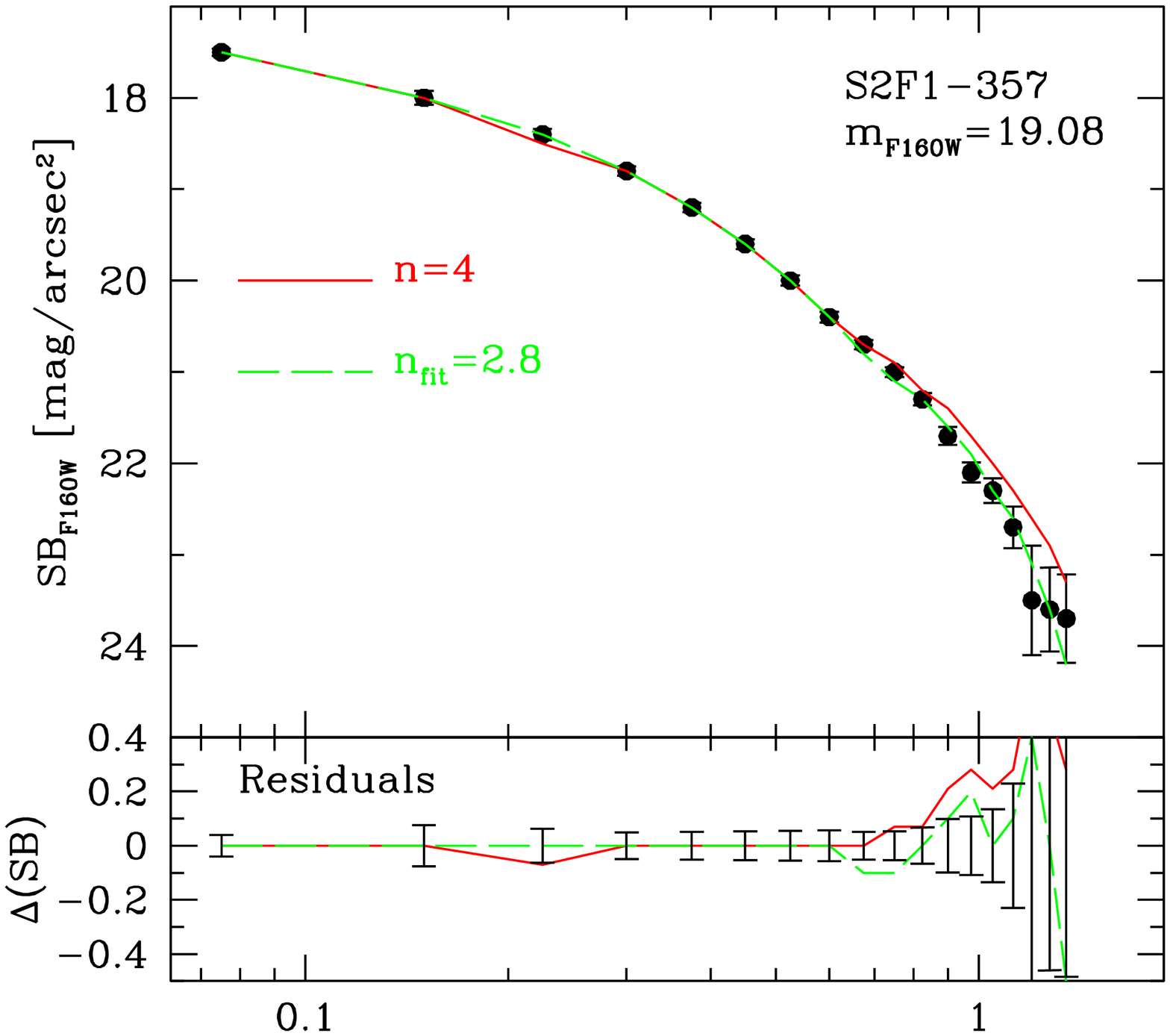}
\vskip -0.5truecm
\includegraphics[width=5.8cm,height=5.8cm]{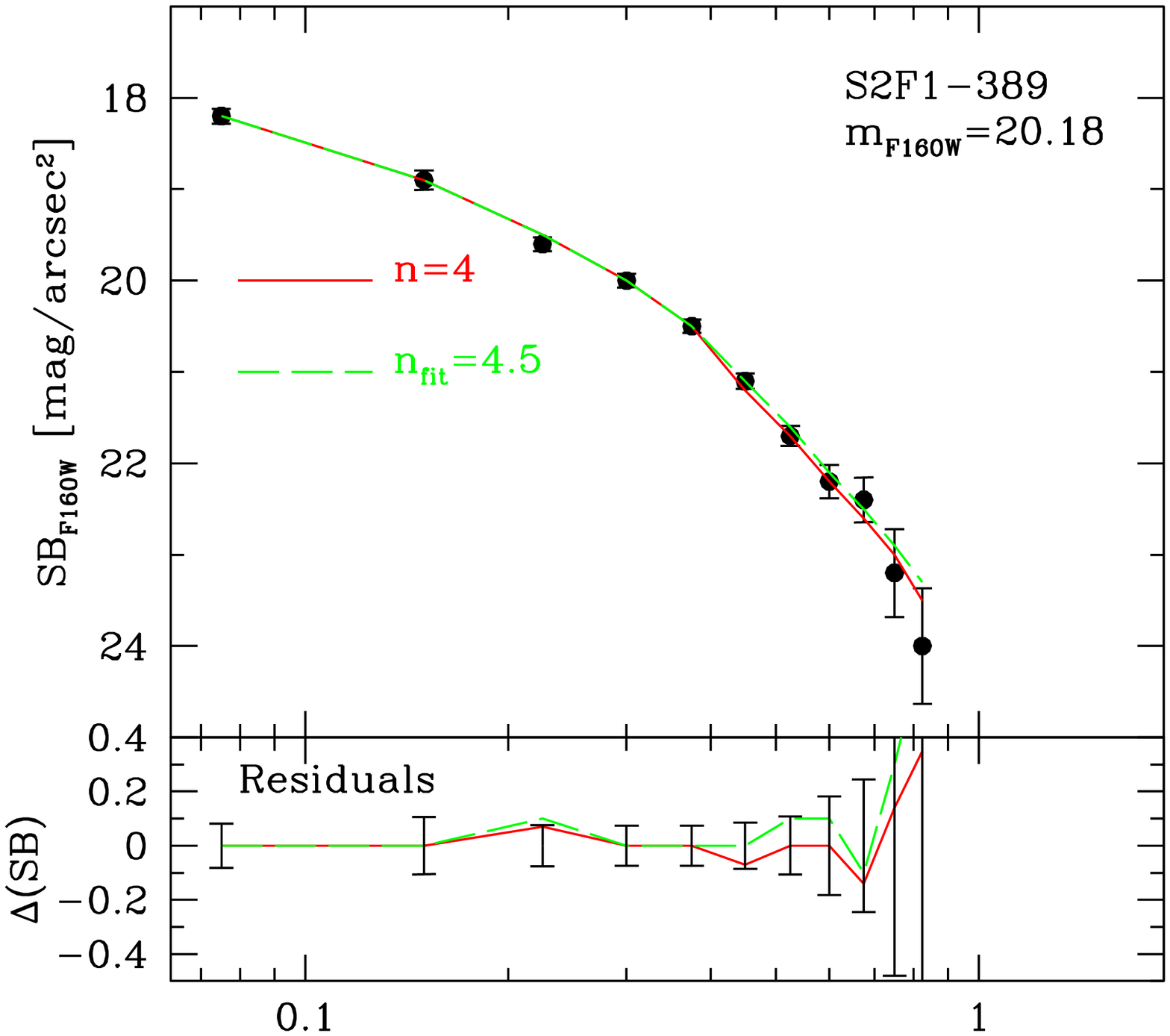}
\includegraphics[width=5.8cm,height=5.8cm]{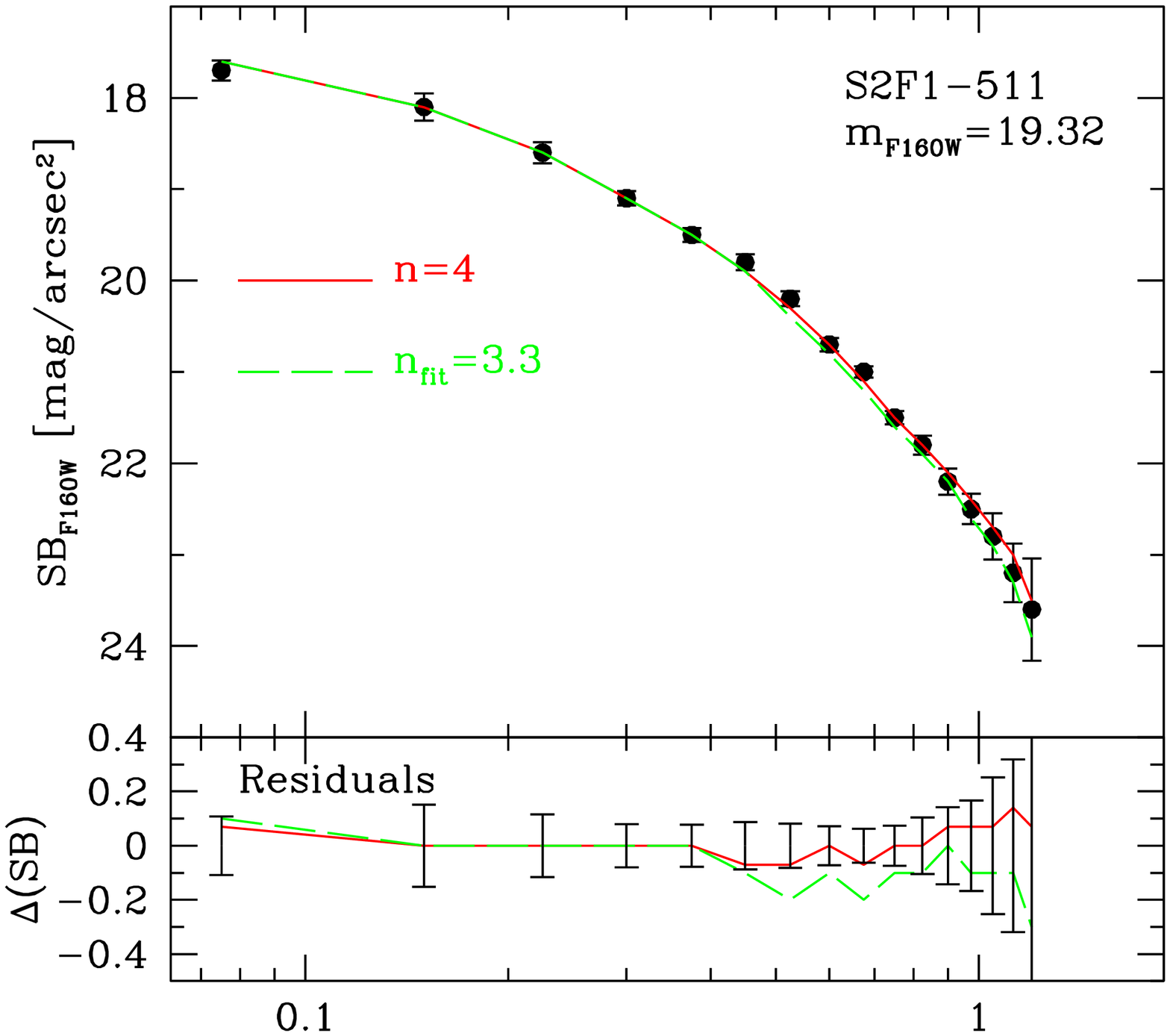}
\includegraphics[width=5.8cm,height=5.8cm]{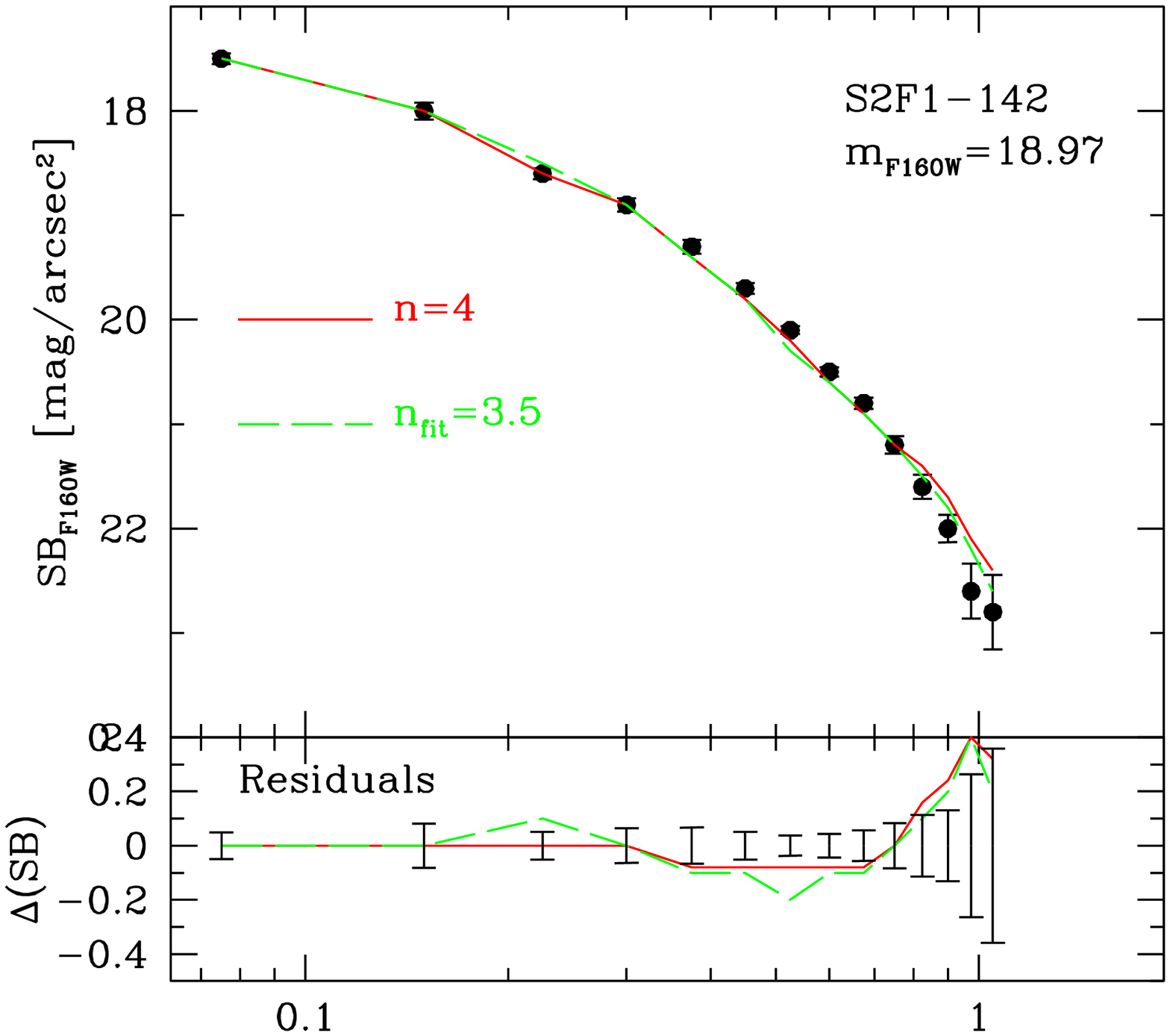}
\vskip -0.5truecm
\includegraphics[width=5.8cm,height=5.8cm]{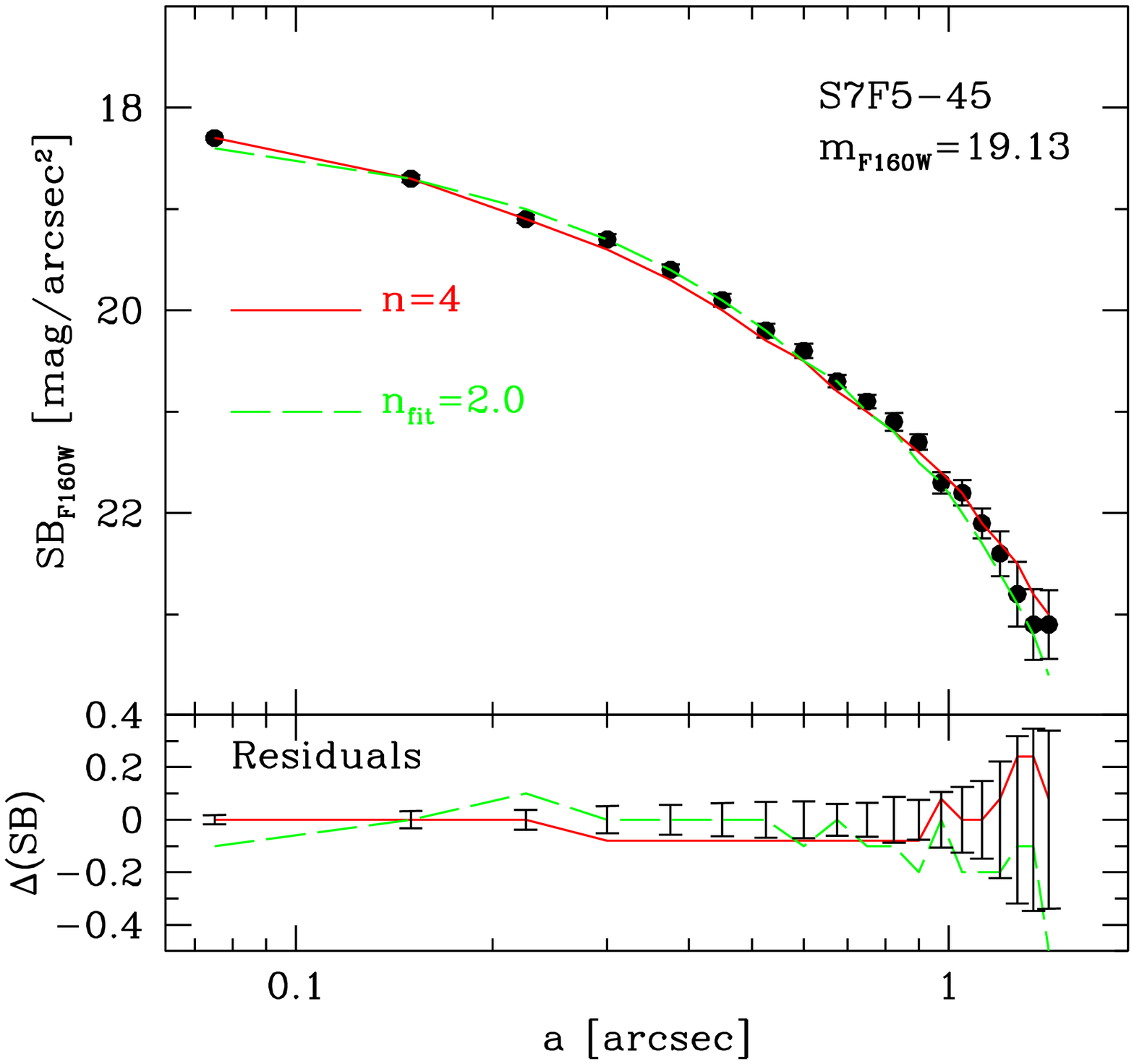}
\includegraphics[width=5.8cm,height=5.8cm]{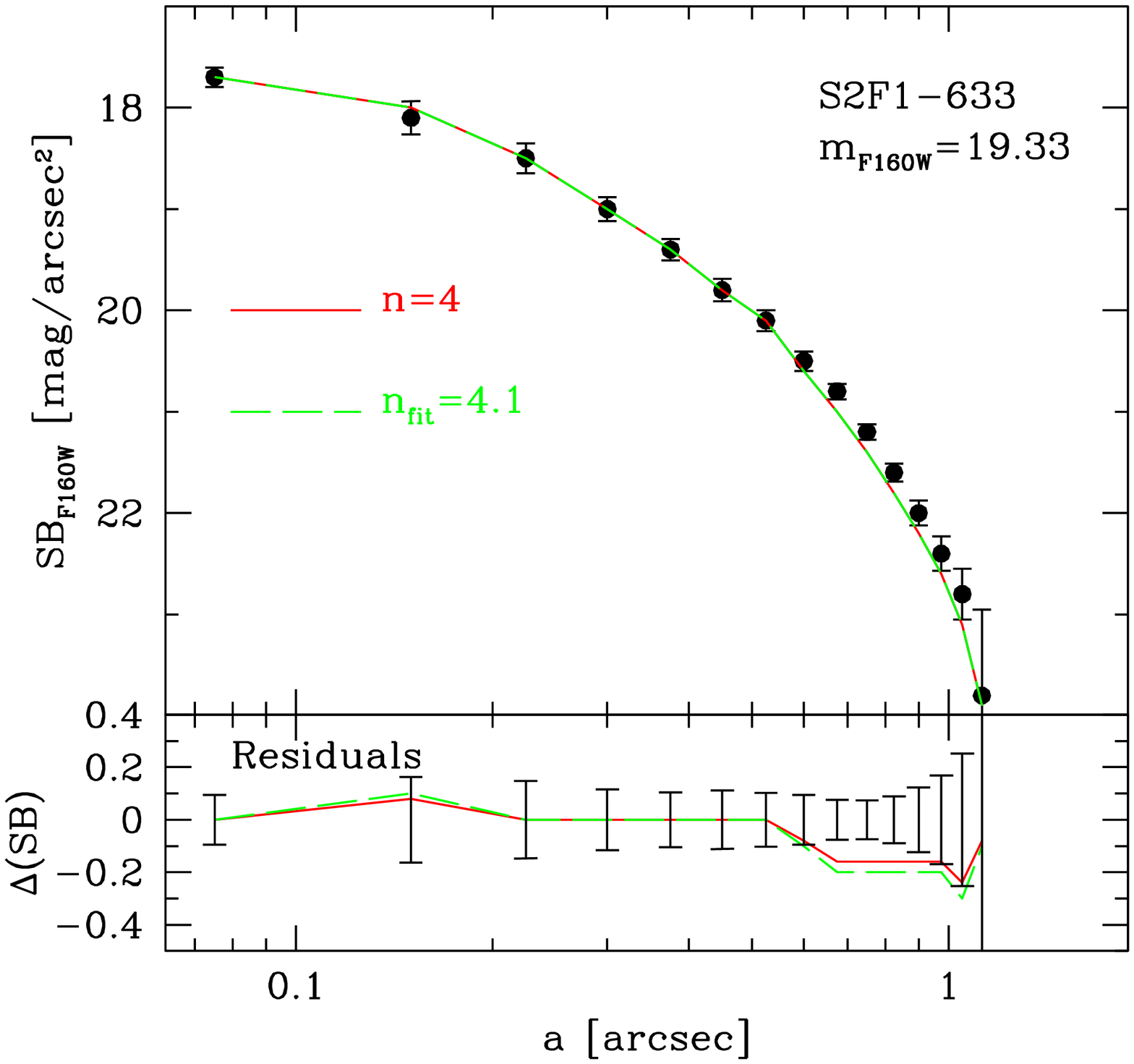}
\includegraphics[width=5.8cm,height=5.8cm]{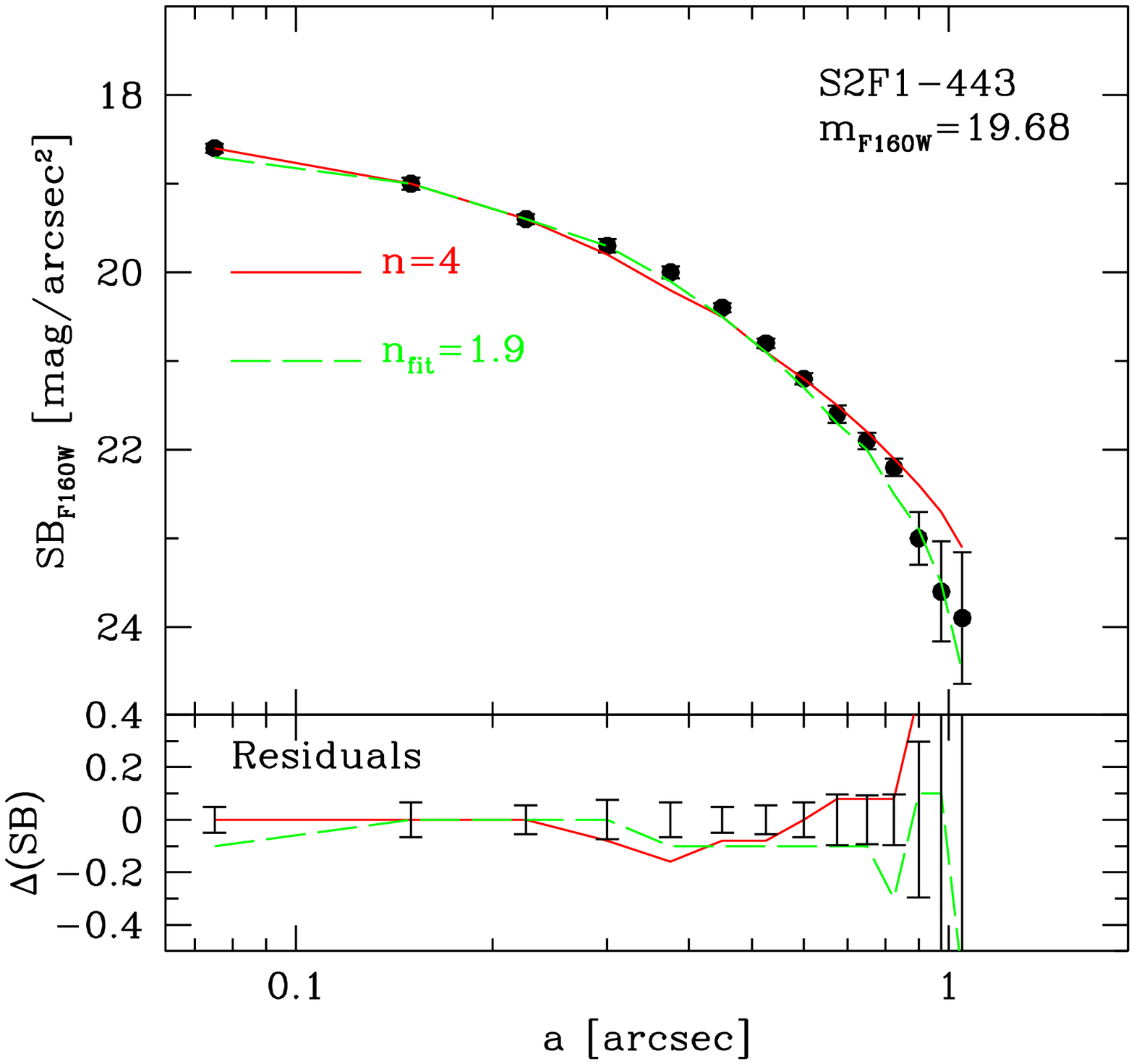}
\caption{In the upper panels of each figure, the  surface brightness 
 in the F160W band measured along the galaxy major axis of each galaxy 
(data points)  is compared with the de Vaucouleurs ($n=4$, solid/red line) 
and the S\'ersic ($n_{fit}$, dashed/green line) profiles resulting from 
the fitting. 
The NIC2 pixel scale is 0.075 arcsec/pixel. 
In the lower panels the residuals of the fitting obtained as the difference 
between the data points and the models are shown. 
The formal 1$\sigma$ errorbar on the data points are shown for comparison. 
}
\end{figure*}

\subsection{Simulations}
To assess the robustness of the results of the
fitting to the profile of our galaxies, 
we applied the same fitting procedure  to a set of simulated galaxies
inserted in the real background.
The main questions  we want to address are $i)$ what is the S{\'e}rsic 
index we would obtain by fitting a S{\'e}rsic profile to a pure $n=4$
profile galaxy if placed at $z\sim1.4$, sampled  
at the angular resolution  of 0.075 arcsec (the pixel size of the NIC2-camera)
and detected with a typical S/N$>20$ and $ii)$ which
is the accuracy in recovering its effective radius.
To this end, we  generated with \texttt{Galfit} a set of 100 galaxies 
described by a  de Vaucouleurs profile  with  axial ratio $b/a$ 
and position angle  PA randomly assigned in the ranges $0.4<b/a<1$ 
and 0$<$PA$<$180 deg respectively.
Magnitudes F160W$_{in}$ and effective radii $r_{e,in}$ have been 
assigned randomly  in the ranges 19.5$<F160W_{in}<$20.5 and $0.2<r_{e,in}<0.5$
arcsec (corresponding to 1.7-4 kpc at $z\simeq1.4$) respectively.
The simulated galaxies have been convolved with the HST-NIC2 PSF  
 described in the previous section  and embedded 
in the real background.
The background image has been constructed by mosaicking
different portions of real images devoid of sources.
The sigma and mask images  have been similarly derived from
the real rms and mask images.
We performed the \texttt{Galfit} fitting
assuming the generalised S{\'e}rsic profile (eq. 1) to the 100  galaxies.
As first result, we find that they  are best fitted by a S{\'e}rsic profile 
with index $2.8<n_{fit}<3.9$ ($<n_{fit}>\simeq3.3\pm0.2$).
It is worth noting that the value of the fitted S\'ersic index  
systematically decreases with fainter magnitudes, i.e. with lower S/N.
This is shown in Fig. 3 where the value of $n_{fit}$
is plotted versus the input magnitude of the simulated galaxies.
{ 
This is possibly due to the fact that lower S/N values
imply the missing of a higher fraction of the tail
of a galaxy profile which would drop abruptly. 
This would favour  lower values of $n$ in the fitting.}

\begin{figure}
\centering
\includegraphics[width=8cm]{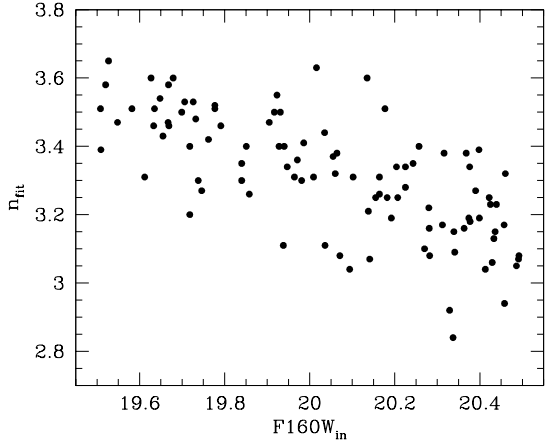}
\caption{Best-fitting S\'ersic index $n_{fit}$ obtained for 
the 100 simulated $n=4$ galaxies  as a function of input magnitude. 
}
\end{figure}

We then  fit the simulated galaxies
with the $n=4$ profile and we studied the behaviour of
the resulting $r_{e,fit}$ with respect to the input $r_{e,in}$ values
for both the two profile fitting functions.
{ In Figure 4 we plot the effective radius $r_{e,fit}$ versus
the input value $r_{e,in}$ in the case of de Vaucouleurs  and
S\'ersic  profiles.} 
In the case of the $n=4$ fitting profile the input effective radius of the 
galaxy is on average underestimated by 
$\Delta r_e^{devauc}\simeq0.025$ arcsec ($\sim10$\%; left 
panel of Fig. 4) while in the case of the S{\'e}rsic fitting profile the  
mean underestimate is $\Delta r_e^{sersic}\simeq 0.07''$ ($\sim25$\%; Fig. 4 
right panel).
It is worth noting that this underestimate is 
smaller than the pixel size  and apparently
independent of  input parameters of the simulated galaxies 
(magnitude, axial ratio, size and PA). 
\begin{figure}
\centering
\includegraphics[width=8cm,height=8cm]{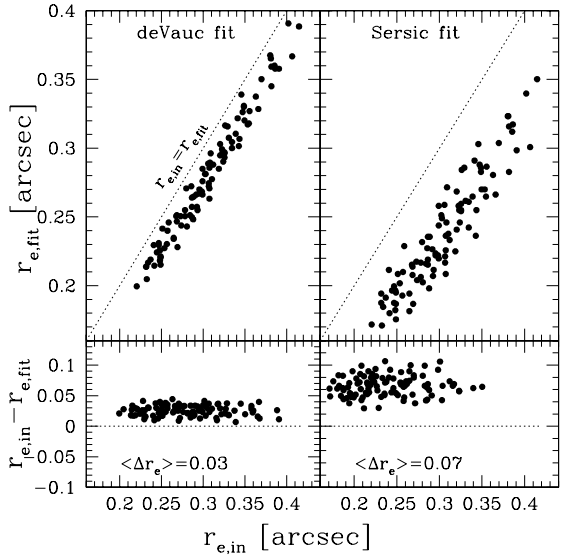}
\caption{Comparison between the effective radius of the simulated galaxies
(r$_{e,in}$) and the effective radius r$_{e,fit}$ obtained through the 
fitting with the de Vaucouleurs profile (left panel) and the S\'ersic 
profile (right panel).}
\end{figure}

{ Finally, we exploited these simulations to assess a reliable magnitude 
estimator for the galaxies in our images.
To this end, we compared the SExtractor MAG\_BEST magnitude and 
the \texttt{Galfit} mag$_{Gal}$ magnitude obtained for the 
100 simulated galaxies to the input $F160W_{in}$ magnitude.
In  Fig. 5 the input $F160W_{in}$ magnitude is plotted versus
the MAG\_BEST magnitude (left panel) and versus the \texttt{Galfit}
magnitude (right panel).
While mag$_{Gal}$ is consistent with $F160W_{in}$,  
MAG\_BEST  is systematically fainter.
The offset is about 0.2 mag with a rather small scatter of $\pm0.03$ mag.
On the basis of this result we used the \texttt{Galfit} 
magnitude as total magnitude of our real galaxies.}

The results obtained from the simulations 
confirm our previous claims that the 6 galaxies with $n_{fit}>2.8$ 
can be considered pure ellipticals.
Only lower values of $n_{fit}$
can be  indicative of a deviation from a pure de Vaucouleurs profile
suggesting the presence of a disk component 
in addition to a dominant bulge.
{ Besides this, we also conclude that \texttt{Galfit} tends to
underestimates the effective radius $r_{e}$ in our images by 
about 0.07 arcsec in the case
of S\'ersic profile and by about 0.025'' in the case of de Vaucouleurs 
profile.
This underestimate has been taken into account in the derivation of the
mean surface brightnesses and will be considered in the following analysis.}
\begin{figure}
\centering
\includegraphics[width=8cm,height=8cm]{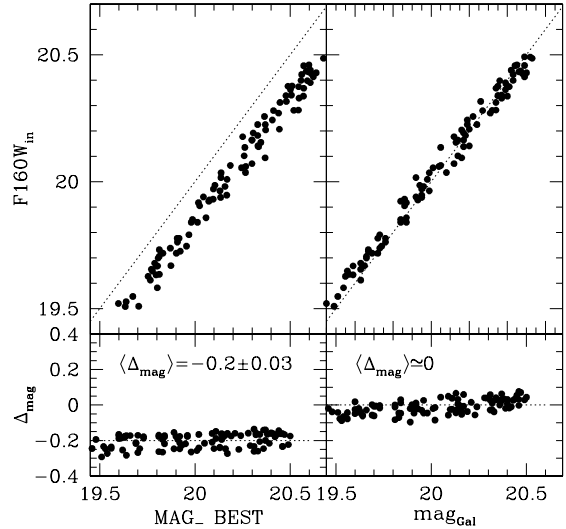}
\caption{Upper panels - Comparison between the SExtractor (left) 
MAG\_BEST magnitude and the  \texttt{Galfit} mag$_{Gal}$ 
magnitude (right) with the input F160W$_{in}$ magnitude of the 100
 simulated galaxies. 
Lower panels - The difference $\Delta_{mag}=F160W_{in}-mag$
is plotted as a function of MAG\_BEST  (left) and of
$mag_{Gal}$ (right).}
\end{figure}

\section{Results}
The first  result of the morphological analysis of
our sample of early-type galaxies is that all of them are bulge 
dominated galaxies.
In particular for 6 of them the best fitting 
S{\'e}rsic profile does not improve the fit with respect to the  
de Vaucouleurs profile providing an index in the range $2.8<n_{fit}<4.5$.
These galaxies can be considered pure elliptical galaxies.
For the remaining 3 galaxies (S7F5\_254, S7F5\_45 and S2F1\_443) 
the values $1.9<n_{fit}<2.3$ fit better the data suggesting
the presence of a disk component possibly contributing up to 30\% 
to the total light of the galaxy. 

A second piece of evidence is that the values of r$_{e}$ obtained when
the S{\'e}rsic light profile is adopted to fit the galaxies light 
distribution are systematically lower than those obtained with the 
de Vaucouleurs light profile. 
This effect was expected also on the basis of the simulations 
showing that both the fitting profiles provide  effective 
radii lower than the input ones.
The underestimate is larger for S{\'e}rsic profile fitting.
{ 
We expect that this underestimate we observe at $z\sim1.4$ would decrease 
with decreasing redshifts  thanks to the higher spatial resolution and 
the larger radii at which galaxies can be sampled.}
This should be taken into account when the local 
morphological relations are compared with those at $z>1$. 

Before doing this comparison, 
we recall the expected behaviour of the two fundamental morphological
parameters used to describe the galaxies, i.e. R$_{e}$ [kpc] and 
$\langle\mu\rangle_{e}$ [mag/arcsec$^2$].
As far as the first one, R$_{e}$, it is usually assumed that it does not evolve
and that it is a characteristic scale of the early-type galaxies depending
only from their mass. 
From this assumption, the expected evolution as function of
redshift of the mean surface brightness within R$_{e}$ can be deduced:

\begin{equation}
\langle\mu_{\lambda}\rangle_{e}=$M$_{\lambda}(z) + 5\ $log$($R$_{e}) + 38.57
\end{equation}

\noindent
where M$_{\lambda}(z)$ is the rest-frame absolute magnitude of the galaxy
as function of $z$ at fixed $\lambda$.
The expected surface brightness evolution of a galaxy, apart from
the cosmological dimming, is due to the luminosity 
evolution expressed by its absolute magnitude at the fixed rest-frame 
wavelength.

In the following section, we will compare the morphological parameters
measured on our sample of early-type galaxies at $z\sim1.4$ with those
of the local early-types. 
The comparison will be carried out on the basis
of the Kormendy scaling relation involving R$_{e}$ and $\langle\mu\rangle_{e}$.

\section{The Kormendy relation}
The Kormendy Relation (Kormendy 1977; KR hereafter) is a linear scaling 
relation between the logarithm of the effective radius R$_{e}$ [kpc] and
the surface brightness $\mu_{e}$ [mag/arcsec$^2$] measured at R$_{e}$: 

\begin{equation}
\mu_{e} = \alpha + \beta \log($R$_{e}) 
\end{equation}

\noindent
The early-type galaxies both in field and in clusters
follow this tight relation with a  fixed slope $\beta \simeq 3$
out to $z\simeq1$ 
(di Serego Alighieri et al. 2005; Reda et al. 2004; 
La Barbera et al. 2003, 2004; 
Ziegler et al. 1999; Schade et al. 1996; Hamabe \& Kormendy 1987).
On the other hand, the zero point $\alpha$ is found to vary
with the redshift  of the galaxies and it reflects
their luminous evolution over the time.
Since the value of $\alpha$ strictly depends on the photometric band
and system selected to derive the morphological parameters, 
its value needs to be transformed into that of a common 
rest-frame wavelength when comparisons at different $z$ are performed.
Furthermore, the KR is sometimes
expressed in terms of the {\it mean} surface brightness $\langle\mu\rangle_{e}$
measured inside R$_{e}$ instead of $\mu_{e}$, and the zero point
of the two formalisms is different: 

\begin{equation}
\langle\mu\rangle_{e} = \alpha' + \beta \log($R$_{e}).
\end{equation}

\noindent
If the projected light distribution for an elliptical galaxy
is well described by the de Vaucouleurs profile, then

\begin{equation}
\alpha'=\alpha - 1.4
\end{equation}

\noindent
and the more general relation between $\alpha$ and $\alpha'$
can be found in Graham \& Colless (1997).
In the following we choose to adopt the KR expressed as in the
formula (5), that is we consider the mean surface brightness
$\langle\mu\rangle_{e}$ measured inside R$_{e}$.
Since our galaxies are at $z\sim1.5$, the 
observed F160W band ($\lambda=16030$\AA) corresponds to
the rest-frame 6400\AA, that is about the rest-frame R-band.
Thus, the comparison of our data with those of galaxies
at lower redshift will be done on a KR derived in the 
standard Cousin R band.
The different k-corrections applied to the various samples
and the way we derived them are described in Appendix A. 
As far as the slope is concerned, we adopt $\beta=2.92\pm0.08$
as reported by the recent work by La Barbera et al. (2003).

The zero point at $z\sim0$ can be derived from La Barbera et al. (2003) 
adopting the $\alpha$ value for the galaxies in the Coma cluster,
$\alpha=18.68$.
Morphological parameters of the galaxies in the Coma cluster
come from J{\o}rgensen, Franx \& Kj{\ae}rgaard (1995) and are measured in
the Gunn {\it r} band. We transform $\alpha(r)$ into $\alpha(R)$ by
adding 0.05 (due to slightly different filters) and -0.43 (due to the
different photometric system of the gunn magnitudes):

\begin{equation}
\langle\mu\rangle_{e} = 18.30 + 2.92 \log($R$_{e}).
\end{equation}

\noindent
If we consider $z_{Coma}=0.024$, for the cosmological dimming we derive
$\alpha=18.20$ at $z=0$.

The zero point at $z\sim0$ can be also derived by the original
Hamabe \& Kormendy V-band relation for early-type galaxies in the local 
universe: 
\begin{equation}
\mu_{e} = 19.48 + 2.92 \log($R$_{e})
\end{equation}

\noindent
Assuming a typical colour (V-R)=0.65 we derive the corresponding R-band
$\alpha=18.83$ 
\noindent
and by means of the equation (6):

\begin{equation}
\langle\mu\rangle_{e} = 17.43 + 2.92 \log($R$_{e}).
\end{equation}
It is worth noting that Hamabe and Kormendy possibly
derive this relation assuming H$_0=50$ km s$^{-1}$ Mpc$^{-1}$, even if 
not explicitly stated.
If this was the case, their relation scaled to H$_0=70$ km s$^{-1}$ 
Mpc$^{-1}$ would agree with the one derived by La Barbera et al.
(eq. 7). 

\begin{figure*}
\centering
\includegraphics[width=16.8cm]{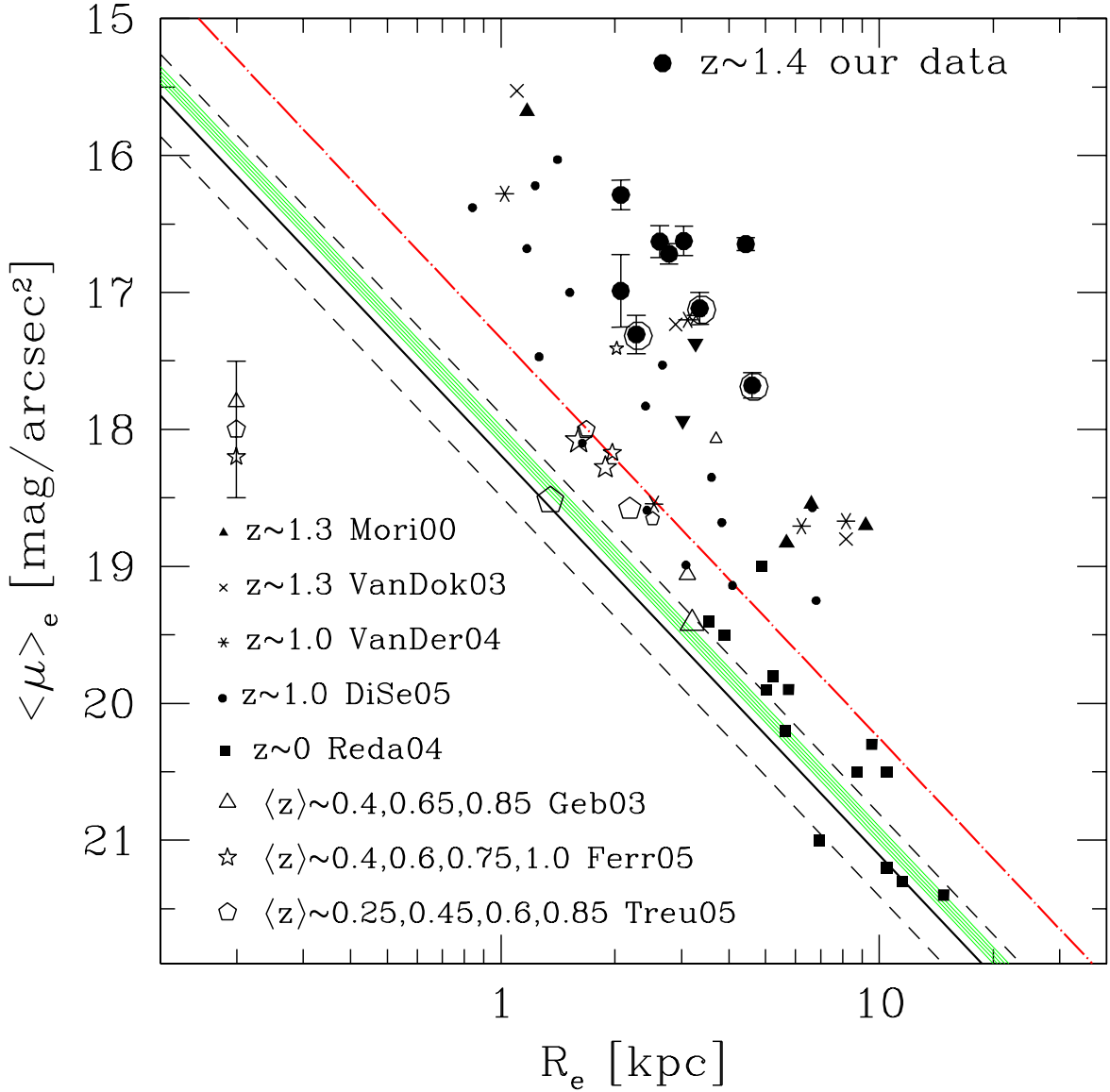}
\caption{Mean surface brightness $\langle\mu\rangle_e$ versus  effective 
radius R$_e$ [kpc].
All the data are corrected for the cosmological dimming $(1+z)^4$.
Our data (large filled circles) are compared with those of other authors.
The large open circle superimposed on the filled one marks the three galaxies 
of our sample with $n_{fit}<2.5$ (S2F1\_254, S2F1\_443 and S7F5\_45). 
The small filled symbols represent the data from Moriondo et al. 
(2000, Mori00), Di Serego Alighieri et al. (2005, DiSe05) and 
Reda et al. (2004, Reda04).
{ The open symbols represent the mean values of the surface brightness of 
field galaxies in the different redshift bins considered as derived from 
the data of Gebhardt et al. (2003, Geb03), Treu et al. (2005, Treu05) and 
Ferreras et al. (2005, Ferr05). 
The size of the symbols decreases with increasing redshift. 
The errorbar on the left marked with the open symbols represent the typical
scatter in surface brightness within each bin.}
The solid line represents the KR at $z\sim0$ found by La Barbera et al. (2003)
as expressed in eq. (7). 
The short-dashed lines represent  the $\pm1\sigma$ dispersion of the relation.
The shaded (green) strip is the KR at $z\sim0$ of eq. (9) (scaled to 
H$_0=70$ km s$^{-1}$) found by  Hamabe \& Kormendy (1987).
The long-dot-dashed (red) line is the KR at $z\sim0.64$ by La Barbera et al. 
(2003).}
\end{figure*}

Fig. 6  report the KR expressed in the equation (7) (solid black line)
and in eq. (9) (shaded green strip), this latter scaled to 
H$_0=70$ km s$^{-1}$ Mpc$^{-1}$. 
The uncertainty of the KR by La Barbera et al. (2003)
is shown as a strip limited by short-dashed black lines and the upper limit
results coincident with the original (uncorrected) KR by Hamabe \& Kormendy 
(1987).  
The grey (red) dot-dashed line has been derived by the KR
reported by La Barbera et al. (2003) for the cluster EIS0048 at $z=0.64$.
In the  figure, our data are reported as large filled circles.
An open circle marks the three galaxies having $n_{fit}<2.5$
(S2F1\_254, S2F1\_443 and S7F5\_45).
In Fig. 6, the $\langle\mu\rangle_{e}$ and R$_{e}$ values of our galaxies
have been measured through the S{\'e}rsic profile fitting. 
The rest-frame surface brightness in the R-band, $\langle\mu\rangle_{e}$,  
has been obtained  from the observed F160W-band apparent magnitudes.  
Since for each of our
galaxies we have a detailed description of its rest-frame optical spectral
shape, we accurately derived the {\it k}-correction:

\vskip 0.1truecm\noindent
{\it k}=mag(F160W;$z=z_{gal}$)-mag(R,$z=0$)
\vskip 0.1truecm

\noindent
where the two magnitudes are derived from the
spectral energy distribution reproducing that of each galaxy.
\footnote{The uncertainty affecting this correction is
much smaller than the photometric errors affecting
the measure of $\langle\mu_{e}\rangle$ itself. 
This is due to the fact that
the observed band at $z=z_{gal}$ is almost coincident with
the rest-frame R band. Moreover, the sample of spectral templates 
used to calculate this {\it k}-correction is constrained to few and similar 
spectra  by the multi-band photometry and by the near-IR spectra available
for all the galaxies (see \S2).}  

In Fig. 6 we also report  data from literature
resulting from both  S\'ersic and de Vaucouleurs 
 profile fitting.
{ Small filled symbols are the values relevant to the galaxies belonging
to the samples of Reda et al. (2004, squares), Di Serego Alighieri et al. 
(2005, circles) and Moriondo et al. (2000, triangles).}
As $z\sim0$ reference, we report the sample of local galaxies
of Reda et al. (2004) and  derived
assuming the de Vaucouleurs profile in the galaxy fitting.
The morphological parameters of the sample of Di Serego Alighieri et al. 
(2005) have been obtained by assuming a general S\'ersic profile.
Their data have been transformed from B-band to R-band (see Appendix A).
The sample of  Moriondo et al. (2000) includes 6 galaxies 
at $z\sim1.3$, 4 out of which have been observed with the same setup of 
our sample of galaxies (HST+NICMOS, F160W filter)
while the remaining 2 have been imaged in the HST+WFPC2 F814 filter
(see Appendix A).
As in the case of Di Serego Alighieri et al. (2005)
morphological parameters have been estimated by assuming general S\'ersic 
profiles.

{ The open symbols represent the median values of the surface brightness of
field galaxies in different redshift intervals as derived from the samples
of Gebhardt et al. (2003, triangles), Treu et al. (2005; pentagon)
and Ferreras et al. (2005; stars).
The size of the symbols decreases with increasing redshift.
The sample of Gebhardt et al. (2003)  includes 
21 early-types with spectroscopic redshift and HST imaging in the 
F814W filter. 
They derive the effective radii  from a de Vaucouleurs fitting 
(Simard et al. 2002).
The sample spans a very narrow range of R$_e$, from 2.9 to 3.9 kpc.
We divided this sample in the three redshift bins $0.28<z<0.5$ (6 gal.)
$0.5<z<0.8$ (7 gal.), $0.8<z<1.0$ (8 gal.) and we derived the surface 
brightness in the rest-frame R-band from the F814W apparent magnitude
(see Appendix A).
The sample of Treu et al. (2005)  includes $\sim90$ galaxies 
with  morphological types 0 (E) and 1 (E/S0),
according to their classification.
The  morphological parameters have been obtained from a de Vaucouleurs profile 
fitting to HST-ACS images.
We considered the four redshift bins $0.1<z<0.3$, $0.3<z<0.5$, $0.5<z<0.7$
and $0.7<z<1.0$ and we derived $\langle\mu_e\rangle^{\rm R}$ from the
apparent magnitude in the F814W.
The sample of Ferreras et al. (2005) includes $\sim70$ galaxies
with spectroscopic redshift in the intervals  
$0.2<z<0.5$, $0.5<z<0.7$, $0.7<z<0.9$ and $0.9<z<1.1$.
We selected them according to the photometric type 
$t_{phot}<1.5$.
We derived the rest-frame R-band surface brightness from the F775W 
apparent magnitude.
It is worth noting that the selection criteria used by Ferreras et al.
to construct this sample are quite different from those used 
for the other samples.
Indeed, they classify the galaxies  on the basis
of the best-fitting template  and then they exclude from the final
sample those galaxies that are not consistent with an evolution into
the Kormendy relation.
}  

Besides these data sets, a morphological study of early-type galaxies at 
$z\sim1$ has been made also by van Dokkum et al. (2003) and by 
van der Wel et al. (2004), represented in Fig. 5 by the starry symbols.
The sample of van Dokkum et al. (2003)
contains three galaxies selected from the cluster RDCS J0848+4453 at $z=1.27$
observed with HST-NICMOS in the F160W filter. 
The sample of van der Wel et al. (2004) is composed by 6 galaxies, 
4 at $z\sim1$ and 2 at $z=0.7$, observed with HST-ACS F850LP and F775W 
respectively. 
It should be noted that the effective radii they report in Tab. 1 are in 
arcsec and not in kpc as stated in the header of their table.  
In both samples the morphological parameters have been derived 
assuming de Vaucouleurs light profiles.

\subsection{Constraining the evolution of early-types galaxies}
Fig. 6 shows that the higher the redshift the more luminous
are the galaxies at fixed R$_{e}$. 
It is worth noting that all the data have been corrected for the cosmological
dimming factor $(1+z)^4$.
Moreover, the rest-frame wavelength sampled  ($\sim 6500$\AA)  is weakly 
affected by episodes of recent star formation while it traces very well
the bulk of the stellar mass of the galaxies. 
Thus, the observed trend qualitatively reflects  the luminosity evolution 
M$_{\lambda}(z)$ that galaxies are expected to undergo with redshift. 
The position of the grey (red) dot-dashed line reveals that the increase 
of $\langle\mu\rangle_e$ and thus of the luminosity  of early-types 
at $z\sim0.6$ is at least 0.5-0.8 mag respect to $z=0$ for the same 
value R$_{e}$.
Approaching $z\sim1$ (see the samples of Di Serego Alighieri et al. 2005, 
Moriondo et al. 2000, Gebhardt et al. 2003 and Ferreras et al. 2005), 
a further increase in luminosity of 0.5-0.8 mag is evident.
Our sample at $z\sim1.4$ suggests a further increase of 
0.5-0.8 mag with respect to the sample at $z\sim1$.
 By comparing our data with the KR of local ellipticals described by eq. 
(7) we derived the mean difference $\langle\Delta\langle\mu\rangle_e\rangle$ 
between the effective surface brightness of our early-type galaxies at 
$z\sim1.4$ and the one of local early-types having the same R$_e$.
We find $\langle\Delta\langle\mu\rangle_e\rangle=2.5\pm0.5$ mag, 
i.e. early-type galaxies at $z\simeq1.4$ would occupy the KR at $z\sim0$ if 
their R-band luminosity decreases by about 2.5 mag at constant R$_e$
in the last 9 Gyrs (from $z\simeq1.4$ to $z\sim0$).
It is also worth noting that, if we consider only the 6 pure elliptical 
galaxies of our sample this estimate would slightly increase since the 
remaining galaxies show slightly smaller $\Delta\langle\mu\rangle_e$.
This value is rather insensitive to the age of the galaxies of our 
sample as we have verified by considering separately the older from the 
younger galaxies.
To verify if the evolution  of $\langle\mu\rangle_e$ observed at
$0<z<1.5$ is consistent with the luminosity evolution expected 
for early-type galaxies, we considered the maximum luminosity dimming 
that a galaxy can experience  in this redshift range.
To this end we modeled a pure passive evolution with a Simple  Stellar 
Population (SSP, Bruzual \& Charlot 2003) 2.5 Gyr old at $z\sim1.5$ 
(the mean age of our sample of early-types, see Tab. 1) and we derived 
the luminosity evolution in the R-band at $z\sim0$.
We estimate that the maximum dimming that a galaxy with a stellar population
2.5 Gyr old can experience in 9 Gyr is about 1.6 mag.
This value is very weakly sensitive to the assumed mean age.
Indeed, it varies by $\pm0.1$ mag for ages $2.5\pm0.5$ Gyr.
Thus, our galaxies show an evolution of $\langle\mu\rangle_e$ which exceeds 
 $\sim$2.5 times ($\sim$1 mag) the one expected in the case of luminosity 
evolution.
This discrepancy cannot be simply recovered.
A possibility is to hypothesize the evolution
of the other quantity affecting $\langle\mu\rangle_e$, i.e. the effective
radius R$_e$.
If this is the case, the effective radius R$_e$ of early-types 
should increase at least by a factor $\sim1.5$ from 
$z\sim1.5$ to $z\sim0$ to account for this discrepancy.  

In the next subsection, we will follow a  quantitative approach to the 
evolution of the KR and of the luminosity of the early-type galaxies as 
function of $z$.

\subsection{The evolution of the zero point of the Kormendy Relation}
It may be useful to recall that the zero 
point $\alpha$ of the KR represents at any redshift the surface brightness
of galaxies whose effective radius is R$_{e}=1$ kpc. 
The decrease (in magnitudes) of $\alpha$ at fixed $\lambda$ with $z$ is a 
direct measure of the increase of  the surface brightness of the galaxies 
with R$_{e}=1$ kpc. Even in the case of a fixed slope of the KR with $z$, 
the increase of $\langle\mu\rangle_{e| R_{e}=1}$ can be caused by two factors: 
1) the increase of the absolute luminosity at fixed wavelength
of galaxies with R$_{e}=1$ kpc assuming that R$_{e}$ does not evolve with $z$;
2) the decrease of R${e}$ with $z$ of galaxies of fixed mass.
It is generally assumed that only the first factor out of the two mentioned 
above is effectively active in the determination of the evolution of 
$\alpha$ in the KR. 
However, we will see that it could be necessary to hypothesize 
also the evolution of R$_{e}$ with $z$.

In Fig. 7, we report the values of $\alpha$ as function of $(1+z)$ measured
in the common rest-frame R band and derived from the samples described
in the previous subsection. 
For all the samples, we assume a fixed slope of
the KR ($\beta=2.92$). 
All the values have been already corrected for
the cosmological dimming $(1+z)^{4}$, and thus they directly describe 
the evolution of the zero point of the KR with $z$ due to the luminosity and/or
scale evolution of the galaxies. 
The  black filled squares at $z\le0.64$ come from La Barbera 
et al. (2003),  the small filled  circle at $z=1$ has been derived from the 
data of Di Serego Alighieri 
et al. (2005), the filled triangle from the sample of
Moriondo et al. (2000).
{ The open symbols come from the samples of Gebhardt et al. 
(2003, triangles), Treu et al. (2005, pentagon) and Ferreras et al. 
(2005, stars).}
The largest filled circles reported at $z=1.4$ represent our results in the 
case of S{\'e}rsic profile fitting (open circle) and de Vaucouleurs profile 
(filled circle). 
In Table 3 the values of $\alpha$ we derived at $z\sim1$ are reported.  
\begin{table}
\center
\caption{Zero point $\alpha$ of the KR derived from samples at $z\sim1$ for 
$\beta=2.92$. 
{ The values in brackets have been derived considering only the 6 
galaxies of our sample with S\'ersic index $n>2.8$.}}
\begin{tabular}{lccc}
\hline
\hline
  Sample  & $\alpha$(deVauc)  & $\alpha$(Ser) \\
  \hline
OUR                 & 15.67(15.35)$\pm$ 0.45     & 15.53(15.37)$\pm$ 0.47  \\
DiSe05              & -----		  & 16.64$\pm$ 0.44 \\
Mori00              & -----		  & 16.09$\pm$ 0.16  \\
DiSe05 + Mori00     & -----		  & 16.50 $\pm$ 0.42 \\
Geb03 $\langle z\rangle\simeq0.85$ & 16.41$\pm$0.4 &---- \\
Treu05 $\langle z\rangle\simeq0.83$ & 17.20$\pm$0.1 &---- \\
Ferr05 $\langle z\rangle\simeq1.0$ & 16.52$\pm$0.1 &---- \\
\hline
\hline
\end{tabular}
\end{table}

\begin{figure}
\centering
\includegraphics[width=8.8cm]{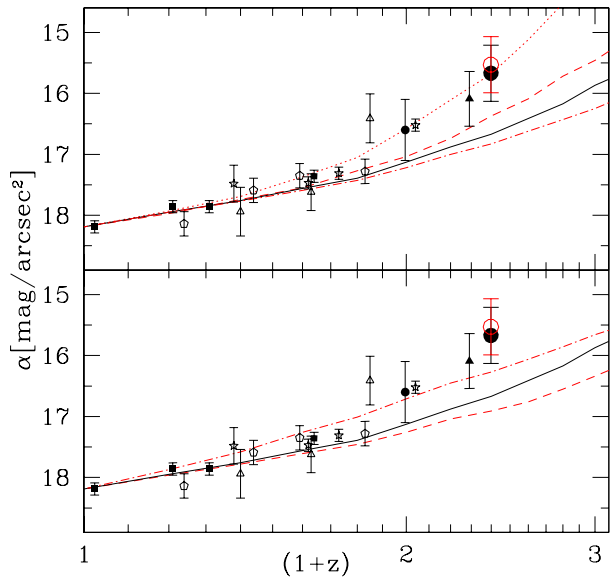}
\caption{Zero points $\alpha$ of the Kormendy relations 
derived by the various samples as a function of their redshift ($\beta=2.92$).
Large filled (empty) circle represent the zero point we obtain from our sample
in the case of de Vaucouleurs (S\'ersic) profile.
The lines show the expected evolution of $\alpha$ due to the luminosity 
evolution
of galaxies, i.e. due to the evolution of their M/L ratio for different models.
Upper panel: models refer to the same SFH $\tau=0.6$ Gyr starting at the
redshifts of formation (from the top to the bottom) $z_f=2, 3, 4, 6$.
Lower panel: models assume the same $z_f=4$ but refer to different SFHs,
from the top, $\tau=2.0, 0.6, 0.1$ Gyr.}
\end{figure}
In Fig. 7, the expected evolution of $\alpha$ due to the
 luminosity evolution of galaxies with fixed mass 
(i.e., due to the decrease of M/L) 
for some models is superimposed on the data.
In the upper panel all the model lines are derived assuming
a fixed star formation history (SFH, $\tau=0.6$ Gyr) at solar metallicity, 
and they differ for the formation redshift $z_{f}=2,3,4,6$ from the top to 
the bottom.
In particular,the  black solid line represents the model with $z_{f}=4$.
In the bottom panel, all the models refer to $z_{f}=4$ and they differ
for the SF history assumed, where $\tau=2.0,0.6,0.1$ Gyr from the top to the 
bottom. 
The black solid line represents the same model as in the upper panel.
All the models are obtained with the BC03 code,
and they assume Chabrier (2003) IMF. 
We verified that the choice of different IMFs does not change the results.

Fig. 7 shows that data at $z \geq 1$ are only marginally compatible
with the expected evolution of the luminosity of the galaxies. Indeed, they 
almost match only lines corresponding to models which include star formation 
activity at $1<z<2$, i.e. $z_{f}\leq2$ for $\tau=0.6$ Gyr (upper panel) or 
$\tau=2$ Gyr for $z_{f}\leq4$ (lower panel). 
On the other hand, the 
spectrophotometric analysis of the whole SED of at least some of these
samples of galaxies, from UV to near IR, shows that the bulk of their 
stellar content must be older than some Gyr. 
Furthermore, many of the samples represented in Fig. 7 have been selected on 
the basis of red optical-near IR colours (e.g., R-K$>$5) which are not 
compatible with stellar contents younger than 1 Gyr on average. 
Thus, if we try to interpret Fig. 7 on the basis
of the luminosity evolution of galaxies we hardly reconcile  models and data.
A possible way to solve this problem is to assume that galaxies evolve
also in their morphological scale, that is to admit that the effective radius
of a galaxy of fixed mass increases from the epoch of its formation toward $z=0$.
As we have already recalled, the zero point $\alpha$ of the KR represents 
the surface brightness of the galaxies with R$_{e}=1$ kpc, and from eq. 2
it is expected to evolve as M$_{\lambda}(z) _{[R_{e}=1]}$ once the 
cosmological dimming has been taken into account: 

\noindent
\begin{equation}
\langle\mu_{\lambda}\rangle_{e |R_{e}=1}\  \approx \  $M$_{\lambda}(z)_{[R_{e}=1]}.
\end{equation}

\noindent
It is usually assumed that M$_{\lambda}(z)_ {[R_{e}=1]}$
varies only as a consequence of the variation
of the stellar mass to light ratio:

\noindent
\noindent
\begin{equation}
$M$_{\lambda}(z)_{[R_{e}=1]}  \approx \ -2.5 $log$[(L_{\lambda}/$M$)(z)]
   -2.5 $log$($Mass$)_{[R_{e}=1]}
\end{equation}

\noindent
where only $L_{\lambda}/$M is function of $z$, while the stellar mass
content of the galaxies with R$_{e}=1$ kpc is constant.
 
\noindent
Assuming evolution of R$_e$ from $z_{f}$ to $z=0$ implies that, 
at each redshift, galaxies of a given R$_e$ have variable mass.
The expected change in the value of $\alpha$ must contain an additive
increase with $z$ due to the increase of mass of the galaxies with fixed scale,
that is Mass$_{[R_{e}=1]}$ becomes Mass$_{[R_{e}=1]}(z)$ function of $z$.
Fig. 8 shows an example of the expected evolution of $\alpha$ with $z$
in the assumption that the evolution of R$_{e}$ stops for $z<0.5$. 
Points and lines represent the same data and models as in Fig. 7, 
but for $z>0.5$ the value of the zero point
of the KR increases by $\Delta(\alpha)=-2.5 \log (0.5+z)$, 
that corresponds to assume:
\begin{equation}
$R$_{e}(z>0.5) ={ {R_{e}(z\le0.5)} \over {0.5+z}}.
\end{equation}
for a fixed stellar mass of a galaxy.
In this case,  the models which describe the spectrophotometric properties 
of the galaxies, 
i.e. $z_{f}\sim4$ and SF histories with short time scales ($\tau<1$Gyr) can
reproduce also the observed behaviour of $\alpha$.
The evolution of the size of the spheroid-like objects has been 
recently hypothesized also by other authors, at least for the most massive 
galaxies. 
For instance, Trujillo et al. (2006b) find strong hints that the most 
massive early-type galaxies were $\sim 4$ times smaller at $z>1.4$ than 
at $z=0$. 
If we consider $z=1.4-1.5$ in eq. (12) we obtain a lower (even
if comparable) evolution. 
It is important to point out that the increase of the stellar mass
at $z>0.5$, needed to explain the observed evolution of the zero point
of the KR, can be hardly ascribed to on-going SF activity at $0.5<z<1.5$. 
Indeed, from one hand the observed SED of the early-type galaxies 
at $z\sim 1.0-1.5$ is not compatible with recent or on-going SF 
(Longhetti et al. 2005) and, on the other hand, our sample at 
$z\simeq1.4$ is composed by galaxies with stellar 
masses M$_{star} \ge 3\times 10^{11}$ M$_\odot$ which cannot form  more stars
at $z<1.4$  to be compatible with the local mass function of
the early-type galaxies. 
Thus, the possible solution is to hypothesise that
R$_{e}$ of early-type galaxies increases from at least $z\sim1.5$ toward
$z=0$ and that this behaviour is not related to a growth
of their stellar mass by star formation.
A possible evolution of the effective radius is suggested also from the
observation that the values of R$_{e}$ measured for the galaxies of
our sample are smaller on average than the values measured
on galaxies at lower redshift (see Fig. 6) in spite of the 
high-mass of our galaxies. 
\begin{figure}
\centering
\includegraphics[width=8.8cm]{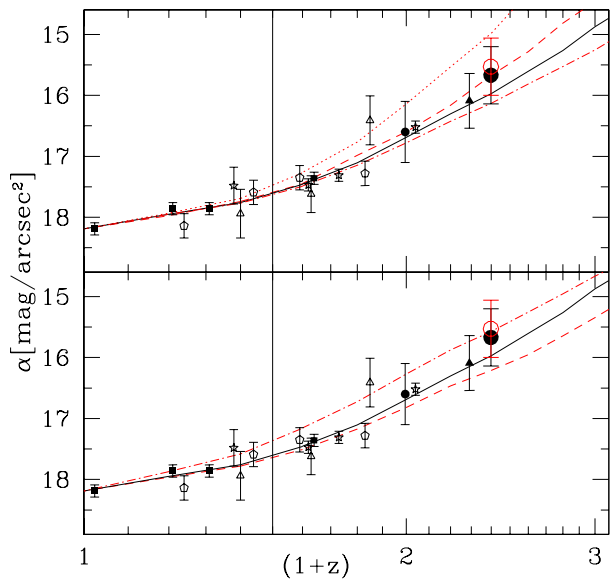}
\caption{Same as Fig. 8 but models include the evolution of the effective radius Re,
as described in \S 5.2.}
\end{figure}
Finally, it is worth  noting that the assumed scale evolution of galaxies 
corresponding to the increase of R$_{e}$ from their formation
toward $z=0$ does not imply any change in the slope of the
KR. It can be seen as single galaxies during their ageing
move {\it along} the KR, causing a further change in its zero point
but not a change in its slope. 

\section{Summary and conclusions}
We presented the morphological analysis of a sample of 9 massive
(M$_{star}> 10^{11} $ M$_\odot$) galaxies spectroscopically classified
as early-types  at $1.2<z<1.7$.
The analysis is based on HST imaging  carried out with the NICMOS-NIC2 
camera (0.075 arcsec/pixel) in the F160W filter ($\lambda\sim16000$ \AA),
sampling the rest-frame R-band. 
We studied their profile and we derived their effective radius r$_{e}$ and 
their mean surface brightness $\langle\mu\rangle_{e}$ within R$_{e}$
by means of the \texttt{Galfit} software (v. 2.0.3; Peng et al. 2002).
From the surface brightness fitting analysis performed we find that
the 9  early-type galaxies analysed are bulge dominated 
in agreement with the spectral classification.
Six of them are very well fitted by the pure de Vaucouleurs profile ($n=4$)
and the addition of the index $n$ as a free parameter 
(providing values $2.8<n_{fit}<4.5$) 
is not statistically required by the fitting.
Thus, these 6 galaxies can be considered pure ellipticals.
For the remaining 3 galaxies the values $1.9<n_{fit}<2.3$ suggest the 
presence of a disk component possibly  contributing up to $30\%$
to the total light of the galaxy suggesting that they are S0 galaxies. 
The median effective radius of our sample of early-types is
$\sim2.95$ kpc and $\sim2.8$ kpc in the case of  de 
Vaucouleurs and S\'ersic profile respectively.
 
By applying the same analysis to a set of simulated galaxies with
intrinsically pure $n=4$ profile embedded in the real background we find 
that they are fitted by values $2.8<n_{fit}<3.9$ confirming the 
previous morphological classification.

The data we have analysed allow us to exclude that the sizes measured
are affected by morphological k-corrections as (possibly) in the
case of blue and UV observations, as suggested by Daddi et al. (2005).
Indeed, our observations sample the rest-frame R-band 
mainly sensitive to the old stellar population.
Moreover, the high resolution (FWHM$\sim0.1$ arcsec) of our HST-NIC2 images,
corresponding to a spatial sampling of $\sim0.8$ kpc at the redshift of
our galaxies,  allow us to exclude  that the estimate of the effective radii
is limited by the resolution.

The analysis of the KR reported to a common rest-frame
R-band of our data compared with other samples at lower redshift
shows that
 the higher the redshift the more luminous are the galaxies 
at fixed R$_{e}$, as qualitatively expected from the luminosity evolution of 
galaxies.
However, we find that the mean surface brightness $\langle\mu\rangle_e$ of 
our sample should evolve by 
about 2.5 mag to $z\sim0$ to occupy the local KR implying a luminosity 
evolution of an order of magnitude in the last 9 Gyr.
This luminosity evolution exceeds by  a factor 2.5 (1 mag) the pure 
passive evolution, the maximum evolution expected for an elliptical galaxy.
This suggests that other parameters, possibly structural, may undergo 
an evolution.

The observed effect mentioned above has been quantified also by means
of the evolution of the zero point $\alpha$ of the KR as function
of $z$.  
The resulting evolution of $\alpha$ cannot be simply explained in terms
of the evolution of the luminosity of the galaxies, at least at $z>0.8$.
Indeed, the strong decrease of $\alpha(z)$ at $z>0.8-1.0$ requires
to assume that the stellar content of the early-type galaxies
at that redshift is too young ($<1$ Gyr) and that they are
still forming stars. 
But these assumptions are not
compatible with their global spectrophotometric properties which
date their stellar content as {\it old} ($\gg1$ Gyr).

A way to reconcile models and observations is to hypothesise that 
morphological scale  plays a role in the evolution of the surface brightness.
We proposed a scheme in which  the effective radius
of a galaxy of fixed mass increases from the epoch of its formation 
toward $z=0$.
In particular, if R$_{e}(z) ={ {R_{e}(z\le0.5)} \over  {(z+0.5)}}$,
the same models describing the spectrophotometric properties of the 
early-types observed at $z>1$ are able to match the observed 
evolution of $\alpha$ with $z$. 
On the other hand, Waddington et al. (2002)
present evidence that the smaller size of their 2 radio galaxies at
$z=1.5$ compared with the sample of more powerful radio galaxies
at $z\sim0.8$ by McLure \& Dunlop (2000) can be explained simply assuming
a lower mass without any need of scale evolution. 
{ It is thus possible that the evolution of R$_{e}$ with $z$ that we have 
proposed involves only galaxies with stellar masses 
M$_{star}>10^{11}$ M$_\odot$. 
The evolution in the galaxy sizes seems to be indeed dependent
on the mass of the objects as pointed out by many authors
(e.g. Treu et al. 2005; Trujillo et al. 2006a, 2006b).
However, the zero points of the KR derived from the
samples of Di Serego Alighieri et al. (2005), Moriondo et al. (2000),
Gebhardt et al. (2003) and Ferreras et al. (2005)
which contain less massive galaxies, present the same inconsistencies 
with models as those derived from our sample, even at a lesser extent.}

Obviously, many problems arises from the assumption of size evolution,
 and questions on the physical  framework in which such an evolution can take 
place.  
{ All our galaxies have high stellar masses (M$_{star}>10^{11}$ M$_\odot$) 
and they account for almost all the number density of local early-type 
galaxies with comparable mass (Saracco et al. 2005).
Thus, not all of them can grow further their mass from $z\sim1.5$ to $z=0$
 to be compatible with the local mass function of early-types,
setting aside the process responsible of the growth.
This implies that mergers (dry or wet) at $z<1.5$ cannot play a major 
role in the evolution of high-mass galaxies in this redshift range,  
at variance with the findings of some authors (e.g. Bell et al. 2006).}
In fact, dry mergers possibly combined with AGN feedback were proposed
to reconcile the observed properties of early-type galaxies, including the 
possible size evolution, with the hierarchical paradigm 
(Khochfar and Burkert 2003; Khochfar and Silk 2006; De Lucia et al. 2006; 
Hopkins et al. 2006).
Dry mergers should occur among gas-poor progenitors of evolved stars
without additional star formation to form massive early-type galaxies.
Since dry mergers are less affected by dissipation mechanism than
mergers involving large amounts of gas (wet mergers), the stellar components 
of the final products are less centrally contracted and, consequently, the
optical sizes of dry mergers products should be larger than the wet mergers
products. 
{ It is difficult to arrange this scenario to our results.
Indeed, since not all of our galaxies can grow further their 
stellar mass,
we can  hypothesize that only a fraction ($\sim$30\%) of them  experience 
dry merger at $z<1.5$ accounting for the local density of 
M$_{star}\sim 10^{12}$  M$_\odot$ galaxies.
However, the remaining fraction of galaxies would still remain without 
a counterpart at $z\sim0$.}

If our galaxies have already completed their growth 
at $z\sim1.5$, dry mergers could  be the mechanism responsible of their
assembly at $z>1.5$ provided that something happens at $z<1.5$ to
enlarge their apparent sizes.
On the contrary, the hypothesis that our early-types are the products 
of wet mergers, which would fit better their small sizes observed 
at $z\sim1.5$, hardly fit their other properties.
Indeed, the old age characterizing their stellar populations and
the relevant formation redshift $z>4$ (Longhetti et al. 2005) 
cannot be accounted for by wet mergers which would induce star formation 
episodes.

Thus, we leave with the evidence in hand that high-mass early-type 
galaxies at $z\sim1.5$ are more compact then their local counterparts.
If this is due to a structural evolution we should consider that
smaller sizes imply higher densities and, consequently, higher velocity 
dispersions.
In particular, since our galaxies are at least 1.5-2 times smaller than those 
at $z\sim0$ they should have velocity dispersions at least 1.2-1.5 times 
larger, i.e. larger than $\sim300$ km/s.  
However, such high-mass and high-density systems are not seen in the local
universe, at least  with this high number density.
Thus, in the hypothesis of a structural evolution something must happens 
at $z<1.5$ to relax these  systems and to match the observed size 
and velocity dispersion of the local population of high-mass early-types. 
On the basis of these considerations we believe that a key observational test
would be the measure of the velocity dispersion of such galaxies and a 
spatial map of their kinematics.
These data would allow us to unambiguously address the question 
whether or not the smaller sizes observed in our massive early-type galaxies 
are due to  higher densities, i.e. to structural and dynamical evolution 
occurring after the completion of their assembly.

\appendix
\section{k-corrections applied to lower redshift samples}
All the  k-corrections calculated throughout this paper are
based on the latest version of the Bruzual \& Charlot models (2003; BC03),
and they are obtained averaging the values which can be derived assuming
the Salpeter (1955) and the Chabrier (2003) Initial Mass Function (IMF).

The original KR reported by La Barbera et al. (2003) for the cluster 
EIS0048 at $z=0.64$ is in the I band, and the corresponding
zero point has been transformed into the common rest-frame R-band
by applying a  k-correction of -0.02 mag  assuming a typical
spectral shape of a template at $z=0.6$ with solar metallicity, 
10 Gyr old, and built by means of an exponentially decaying SFR with 
a time scale $\tau=$0.6 Gyr.

No corrections have been applied to the Reda et al. (2004) data since
they provide the measured quantities in the R-band.

Di Serego Alighieri et al. (2005)  provide measure of r$_{e}$ and 
$\langle\mu\rangle_{e}$ in the B band.
We transformed their morphological parameters in our common rest-frame R-band
by applying an average k-correction of -1.4, that corresponds to $z\sim1$
templates at solar metallicity, 4 Gyr old and obtained assuming $\tau=0.6$ Gyr.

The sample of 6 galaxies at $z\sim1.3$ by Moriondo, Cimatti \& Daddi (2000)
is composed of 4 galaxies observed 
with the same setup of our sample of galaxies (HST+NICMOS, F160W filter)
and 2 galaxies in the HST+WFPC2 F814 filter.
The values of $\mu_{ABmag}$ have been transformed into our standard
common R-band rest-frame by applying k-corrections of +0.7 and -1.7
for the F160W filter and F814 filter respectively ($z\sim1.3$ template,
at solar metallicity, 4 Gyr old, $\tau=0.6$ Gyr). 
Another further correction applied to this set of data is due 
to the fact that the authors compute  $\mu_{e}$ instead of $\langle\mu\rangle_{e}$. 
We adopt the Grahams \& Colless (1997) relation to obtain this correction 
factor. 

{ Gebhardt et al. (2003) provide measure of R$_e$,  of the absolute
magnitude M$_B$ and of the $\langle\mu\rangle_{e}$ in the rest-frame 
B band in their Tab. 1.
We divided the sample in three redshift intervals with median redshift
0.4, 0.63 and 0.85.
For each galaxy we derived  $\langle\mu\rangle_{e}$ in the R-band
rest-frame using the 
apparent magnitude F814W and applying a k-correction of 
0.35 at $z\sim0.4$, 0.15  at $z\sim0.63$ and -0.25  at $z\sim0.85$.
We considered  templates at solar metallicity with $\tau=0.6$ Gyr
with age 8, 7 and 5 Gyr respectively.

Treu et al. (2005) provide a sample of more than 220 galaxies with 
morphological type at $0.2<z<1.2$.
We selected those having morphological type 0 (E) and 1 
(E/S0) and divide them into four redshift intervals centered at the median
redshift 0.25, 0.45, 0.6 and 0.85.
We derived $\langle\mu\rangle_{e}$ in the rest-frame R-band from the
photometry in the F814W band.
Also in this case, we applied to each galaxy a k-correction estimated at  
the median redshift of the relevant interval, namely 0.5, 0.3, 0.17
and -0.25 mag derived by  templates 10, 8, 7 and 5 Gyr old respectively
(solar metallicity and $\tau=0.6$ Gyr).

The sample of Ferreras et al. (2005) has been distributed in the
four redshift interval centered at 0.38, 0.62, 0.74 and 1.0.
We derived $\langle\mu\rangle_{e}$ in the rest-frame R-band from the
photometry in the F775W band.
The k-corrections relevant to the redshift intervals are
0.35, 0.15, -0.04 -0.6 mag.
}

The sample of van Dokkum et al. (2003)
contains three galaxies selected from the cluster RDCS J0848+4453 at $z=1.27$
observed with HST-NICMOS in the F160W filter. 
The k-corrections to transform their data into
the standard rest-frame R-band are provided by the authors.

The sample of van der Wel et al. (2004) is composed by 6 galaxies, 
4 at $z\sim1$ and 2 at $z=0.7$, observed with HST-ACS F850LP and F775W 
respectively. 
Small corrections are needed to be applied to the photometric data of 
this sample to be transformed in the standard rest-frame R-band 
(i.e., +0.05, corresponding to the average value derived from templates 
at the 2 different redshift, at solar metallicity, with age between 4 and 
5.5 Gyr, obtained assuming $\tau=0.6$ Gyr). 
For both the samples, the surface brightness provided by the
authors is $\mu_{e}$ instead of $\langle\mu\rangle_{e}$. 
Since for both the samples the morphological parameters have been derived 
assuming de Vaucouleurs light profiles, we transformed $\mu_{e}$ into 
$\langle\mu\rangle_{e}$ by means of eq. (6).

\section*{Acknowledgments}
Based on observations made with the NASA/ESA Hubble Space Telescope, 
obtained at the Space Telescope Science Institute, which is operated by the 
Association of Universities for Research in Astronomy, Inc., under NASA 
contract NAS 5-26555. 
This work has received partial financial support from the Istituto Nazionale
di Astrofisica (Prin-INAF CRA2006 1.06.08.04).
We would like to thank Ignacio Trujillo for the helpful scientific discussions,
for the many suggestions and for a careful reading of the manuscript.
We thank the anonymous referee for the useful and constructive comments.

\label{lastpage}

\end{document}